%% file: tqe.tex
\documentclass{ieeeaccess}
\input{stylesNdefinitions}
\def\BibTeX{{\rm B\kern-.05em{\sc i\kern-.025em b}\kern-.08em
    T\kern-.1667em\lower.7ex\hbox{E}\kern-.125emX}}
\begin{document}
\history{Date of publication xxxx 00, 0000, date of current version xxxx 00, 0000.}
\doi{}

\title{An Analysis of the Completion Time of the BB84 Protocol}
\author{\uppercase{Sounak Kar}\authorrefmark{1} 
\uppercase{and Jean-Yves Le Boudec\authorrefmark{1}}, \IEEEmembership{Fellow, IEEE}}
\address[1]{EPFL, Switzerland (email: firstname.lastname@epfl.ch)}

\markboth
{Kar \headeretal: An Analysis of the Completion Time of the BB84 Protocol}
{Kar \headeretal: An Analysis of the Completion Time of the BB84 Protocol}


\begin{abstract}
The BB84 quantum key distribution (QKD) protocol is based on the idea that the sender and the receiver can reconcile a certain fraction of the teleported qubits to detect eavesdropping or noise and decode the rest to use as a private key.
Under the present hardware infrastructure, decoherence of quantum states poses a significant challenge to performing perfect or efficient teleportation, meaning that a teleportation-based protocol must be run multiple times to observe success.
Thus, performance analyses of such protocols usually consider the completion time, i.e., the time until success, rather than the duration of a single attempt.
Moreover, due to decoherence, the success of an attempt is in general dependent on the duration of individual phases of that attempt, as quantum states must wait in memory while the success or failure of a generation phase is communicated to the relevant parties.
In this work, we do a performance analysis of the completion time of the BB84 protocol in a setting where the sender and the receiver are connected via a single quantum repeater and the only quantum channel between them does not see any adversarial attack.
Assuming certain distributional forms for the generation and communication phases of teleportation, we provide a method to compute the moment generating function (MGF) of the completion time and subsequently derive an estimate of the CDF and a bound on the tail probability.
This result helps us gauge the (tail) behaviour of the completion time in terms of the parameters characterising the elementary phases of teleportation, without having to run the protocol multiple times.
We also provide an efficient simulation scheme to generate the completion time, which relies on expressing the completion time in terms of aggregated teleportation times.
We numerically compare our approach with a full-scale simulation and observe good agreement between them.
\end{abstract}



\titlepgskip=-15pt

\maketitle

\input{sections/intro}
\input{sections/bb84duration}
\input{sections/methodsNproofs}
\input{sections/sytheticSimulation}
\input{sections/eval}
\input{sections/conclusion}
\input{sections/appendix}
\input{sections/bib}
\EOD
\end{document}

%% file: stylesNdefinitions.tex
\usepackage{graphicx}
\usepackage[T1]{fontenc}
\usepackage{hyperref}
\usepackage{enumitem}
\usepackage{widetext}
\usepackage{bbm}
\usepackage{amsmath,amssymb,amsfonts}
\usepackage{cleveref}
\usepackage{geometry}
\usepackage{marginnote}
\usepackage{xcolor}
\usepackage{amsthm}
\usepackage{subcaption}
\usepackage{braket}
\usepackage[ruled]{algorithm2e}
\usepackage{caption}
\usepackage{mathtools}
\usepackage{cite}
\usepackage{algorithmic}
\usepackage{textcomp}

\newtheorem{theorem}{Theorem}
\newtheorem{prop}{Proposition}
\newtheorem{lemma}{Lemma}

\newtheorem{corollary}{Corollary}

\theoremstyle{remark}

\theoremstyle{definition}

\theoremstyle{nonumberbreak}

\newcommand{\ceil}[1]{\left\lceil{#1}\right\rceil}

\newcommand{\brac}[1]{\left({#1}\right)}
\newcommand{\expect}[1]{\mathsf{E}\!\left({#1}\right)}

\newcommand{\prob}[1]{\mathsf{P}\brac{#1}}
\newcommand{\indicator}[1]{\mathbbm{1}_{#1}}

%% file: sections/intro.tex
\section{Introduction}\label{sec:introduction}

Teleportation facilitates the communication of quantum information between nodes separated by physical distance.
Unlike classical communication, the available hardware for performing teleportation is not efficient, which means that multiple attempts must be made to perform a single successful teleportation.
It is therefore more meaningful to consider the completion time, i.e. the time to success, of a teleportation than the duration of a single attempt.
Furthermore, the duration of each phase of a teleportation attempt determines the degree of decoherence of the qubits involved, and hence the success of the attempt.
This phenomenon is a common theme in quantum communication~\cite{vanMeterBook}, where attempts are repeated and their success generally depends on the specific details of each attempt.
The authors in~\cite{brand, liRus, coopmans2022improved} have modelled this behaviour in a discrete-time setting to analyse the completion time of certain quantum communication modules. \looseness = -1

In this work, we extend their approach to a continuous-time setting and do a performance analysis of the completion time of the BB84 protocol~\cite{bb84qkd}.
Our main focus is on the execution of the protocol under realistic hardware infrastructure, rather than considering its security aspects.
Specifically, we assume that the users are connected via a single first-generation repeater~\cite{munroRepeaters}, i.e., a repeater that does not rely on quantum error correction, and the quantum channel between them does not experience any attack.
In such a benign setting, the BB84 protocol can fail due to inefficient hardware or decoherence of quantum states, which occurs, for example, during the communication phases of the protocol.
In our formulation, we consider the local operation and classical communication (LOCC) phases to have positive durations to account for the effect of decoherence during these intervals. \looseness = -1

Our first contribution is to provide a formula for the MGF of the completion time where the relevant phases last for a non-negligible random time.
The MGF can then be used to obtain the Laplace transform of the \textit{CDF} of the completion time, which can be subsequently inverted to derive the CDF.
We also derive a delay bound on the tail probability using Chernoff's method.
This result helps us understand the behaviour of the completion time with respect to the parameters of the elementary stages of teleportation without having to go through the tedious exercise of running the protocol multiple times.
Note that the completion time in this framework also provides a benchmark for the completion time in the presence of an eavesdropper, with other factors remaining the same. \looseness = -1

Our second contribution is a fast simulation scheme to generate the completion time of the protocol.
There has been a considerable amount of work~\cite{qiaoSim,sahooSim,halipSim,jasimSim,minaScalableSim} on the simulation of the BB84 protocol with or without attack, where the focus has been on obtaining the key rate or the eavesdropping detection probability.
All of these works simulate the individual phases of the protocol to arrive at their end objective.
In this work, we demonstrate that this is not necessary if we only want to obtain the completion time, and we assume that the durations of the relevant phases follow certain distributions.
Under such assumptions, we first show that the total time taken to teleport a qubit can be well-approximated by a Coxian phase-type distribution, which can be efficiently aggregated to simulate the completion time of the BB84 protocol. \looseness = -1


The rest of the paper is structured as follows: we describe the setup and main results on the BB84 protocol in Sect.~\ref{sec:bb84duration}, whereas the methods and proofs are given in Sect.~\ref{sec:methods}. 
We next describe the simulation scheme in Sect.~\ref{sec:synSim}, 
while a numerical evaluation of the derived results is given in Sect.~\ref{sec:eval}.

%% file: sections/bb84duration.tex
\section{Assumptions and Main Results on the BB84 Protocol}\label{sec:bb84duration}

\subsection{Setup and Assumptions}\label{sec:bb84assumptions}

We begin this section by recalling the characteristic features of the BB84 protocol~\cite{bb84qkd}.
In this protocol, the sender Alice encodes each bit of a binary sequence of length $n$ using one of two bases chosen uniformly at random.
Once the basis is chosen, $0$'s and $1$'s are coded using orthogonal states in that basis.
For example, Alice may encode $\text{Bin}(n,1/2)$ bits using photon polarisation in horizontal-vertical basis, with H-polarised photons ($\ket{0}$) representing $0$ and V-polarised ones ($\ket{1}$) signifying $1$, while encoding the rest of the bits in diagonal polarisation basis, with D-polarised ($\ket{+}$) and A-polarised ($\ket{-}$) photons representing $0$ and $1$, respectively.
After receiving the encoded qubits, Bob measures them in one of these bases, chosen uniformly at random.
Alice and Bob then announce their bases for each bit and keep the ones where the bases agree, which occurs for $\text{Bin}(n,1/2)$ bits.
Out of these bits, a random sample is chosen, and the protocol is said to be successful if the bits match for a certain fraction of the sample.
The remaining bits are then used as a key.
To analyse the protocol, we further assume the following: 

\begin{enumerate}[label={A\arabic*}]

\item \label{A1:singleChannel} \textbf{Single channel:
} Alice and Bob are connected by a single quantum channel, implying that the qubits are teleported in sequence.

\item \label{A2:singleRepeater} \textbf{Single repeater:} Alice and Bob are just far enough apart to be connected via a single quantum repeater.

\item \label{A3:noAttack} \textbf{No eavesdropping:} There is no eavesdropping on the channel while the qubits containing the key data are teleported.

\item \label{A4:linkModelling} \textbf{Link modelling:} Following~\cite{liRus}, we model an elementary link between a node and the repeater as Werner state~\cite{werner}.
A Werner state with Werner parameter $w$ is given by:
\begin{align*}
  &\frac{1+3w}{4}\ket{\Phi^+}\!\bra{\Phi^+}+\frac{1-w}{4}\bigg(\ket{\Phi^-}\!\bra{\Phi^-} \\
  & \quad +\ket{\Psi^+}\!\bra{\Psi^+}+\ket{\Psi^-}\!\bra{\Psi^-}\bigg)~,  
\end{align*}
where $\ket{\Phi^+}, \ket{\Phi^-}, \ket{\Psi^+},$ and $\ket{\Psi^-}$ denote the Bell states.
If not used immediately, the effect of decoherence on the link is captured by the following formula:
\begin{align}\label{eq:wernerDecay}
   w(t) = w_0 e^{-t/t_c}~, 
\end{align}
where $w(t)$ is the Werner parameter of the link after time $t$ from generation, $w_0$ is the Werner parameter of a freshly generated link, and $t_c$ denotes the joint coherence time of the involved memories.
Following a Bell state measurement (BSM) at the repeater with two elementary links having Werner parameters $w_A$ and $w_B$, an end-to-end link is generated with Werner parameter $w_A \cdot w_B$.

\item \label{A5:noiseModel} \textbf{Single qubit decoherence:}  To account for the decoherence of a single qubit while waiting in memory, we adopt from~\cite{nielsenChuang,vardoyanPerf} the dephasing and asymmetric amplitude damping noise model:
\begin{align}\label{eq:noiseModel}
    &\mathcal{N}_t(\rho) \!= (1\!-\!p(t))(M_0 \rho M_0^\dagger\!+\!M_1 \rho M_1^\dagger) \nonumber \\
    & \thickspace \!+\! p(t) Z(M_0 \rho M_0^\dagger\!+\!M_1 \rho M_1^\dagger)Z,
\end{align}
where the density matrix $\rho$ (resp. $\mathcal{N}_t(\rho)$ ) represents the state of the system at time $0$ (resp. time $t$), $p(t) = (1-e^{-t/t_{de}})/2$ and
\begin{align*}
    M_0 \!=\! \begin{bmatrix}
    1 & 0 \\
    0 & \sqrt{1\!-\!\gamma(t)}
    \end{bmatrix}, \thickspace
    M_1 \!=\! \begin{bmatrix}
    0 & \sqrt{\gamma(t)} \\
    0 & 0
    \end{bmatrix}, 
\end{align*}
with $\gamma(t) = 1-e^{-t/t_{da}}$.
Here, $t_{da}$ and $t_{de}$ are two constants characterising the amplitude damping and the dephasing effect on the concerned memory, respectively.

\item \label{A6:durationsSE} \textbf{Individual phase durations:} We assume that the durations of the atomic phases of teleportation (which we describe next) and that of the reconciliation phase  (where Alice and Bob check if the protocol was successful) follow shifted exponential distribution with parameters specified in Tab.~\ref{tab:notation}.
\end{enumerate}

Under assumption~\ref{A2:singleRepeater}, the teleportation process is executed in the following order:
\begin{itemize}
\item Phase \texttt{LINK-GEN}: Link-level entanglement is established between a user (Alice/Bob) and the repeater. 
\item Phase \texttt{L-COMM}: Immediately after, success/failure of the \texttt{LINK-GEN} phase is communicated to the user and the repeater. 
Note that \texttt{LINK-GEN} and \texttt{L-COMM} run in parallel for Alice and Bob.
\item Phase \texttt{S-COMM}: As soon as both links are successfully generated, a BSM is performed at the repeater which, if successful, results in end-to-end entanglement between Alice and Bob. A direct failure of the measurement operation is observed with probability $(1-p_{\text{swap}})$. Otherwise, the measurement result is communicated to Alice and Bob. 
\item Phase \texttt{T-COMM}: Once Alice and Bob share an entangled resource, teleportation starts immediately. As part of the process, Alice sends measurement results to Bob, who accordingly applies a unitary operation to his qubit. \looseness=-1 
\end{itemize}

\begin{figure}
    \centering
\makebox[0.99\columnwidth]{\fbox{
\begin{minipage}{0.99\columnwidth}
  {\vspace{1pt}
  \small{
  \noindent V0: a simpler version of the BB84 protocol where Bob measures all qubits in the same basis as Alice encoded and uses all of them for reconciliation,\newline
  V1: the actual version of the BB84 protocol,\newline
  $n$: number of qubits teleported in a BB84 attempt, \newline
  $\alpha$: the fraction of the sampled qubits out of the ones where Alice and Bob's bases match, \newline
  $\beta$: the threshold fraction for which measurements should agree for the protocol to be successful, \newline
  $M_Z(t)= \expect{e^{tZ}}$ for an RV $Z$, \newline   
  $W_{n,c}$: completion time of version V0,\newline
  $W_{n}$: completion time of version V1,\newline
  $T_{GA}$ (resp. $T_{GB}$): the duration of the phase \texttt{LINK-GEN} between Alice (resp. Bob) and the repeater, distributed as IID $\text{SE}(\lambda_{\text{gen}},a_{\text{gen}})$, \newline
  $p_{\text{gen}}$: link generation attempt success probability,\newline
  $T_{CA}$ (resp. $T_{CB}$): the duration of the phase \texttt{L-COMM} between Alice (resp. Bob) and the repeater, distributed as IID $\text{SE}(\lambda_{\text{com}},a_{\text{com}})$, \newline
  $p_{\text{swap}}$: BSM success probability at the repeater,\newline
  $T_C^{'}$: the duration of the phase \texttt{S-COMM}, distributed as $\text{SE}(\lambda_{\text{swap}},a_{\text{swap}})$,\newline
  $w_0$ (resp. $w$): the Werner parameter of a freshly generated (resp. at a general point in time) link,\newline
  $t_c$: the joint coherence time of memories involved in BSM at the repeater,\newline
  $T_C^{''}$: the duration of the phase \texttt{T-COMM}, distributed as $\text{SE}(\lambda_{\text{AB}},a_{\text{AB}})$,\newline
  $t_{de}$ (resp. $t_{da}$): the dephasing (resp. amplitude damping) constant characterising the concerned quantum memory,\newline
  $K_C$: the duration of the reconciliation phase \texttt{K-COMM}, distributed as $\text{SE}(\lambda_{\text{AB}},a_{\text{AB}})$, \newline
  $T_{HA}$ (resp. $T_{HB}$): the total time until Alice (resp. Bob) detects the success of link level entanglement generation, \newline
  $V_A = T_{HA}+T_{CA}$, \newline
  $V_B = T_{HB}+T_{CB}$, \newline
  $T_\gamma$: the total duration of failed swap trials in a teleportation attempt, \newline
  $X$: the duration of a single teleportation attempt, i.e., $X = T_\gamma+\max\{V_A,V_B\}+T_C^{'}+T_C^{''}$~,\newline
  $Y$: the indicator variable denoting that Bob's measurement result for a qubit coincides with Alice's.
  }
  }
\captionof{table}{Primary notations. 
}
\label{tab:notation}
\end{minipage}
}}
\end{figure}

Recall that once all qubits have been teleported, Alice and Bob reconcile their measurements for a fraction of the qubits and decide if the protocol is successful.
We denote the corresponding phase as \texttt{K-COMM}.
\textit{Note that if there is a failure at any intermediate phase, the protocol restarts immediately.}

\begin{figure*}[t!]
    \centering \includegraphics[width=0.9\linewidth]{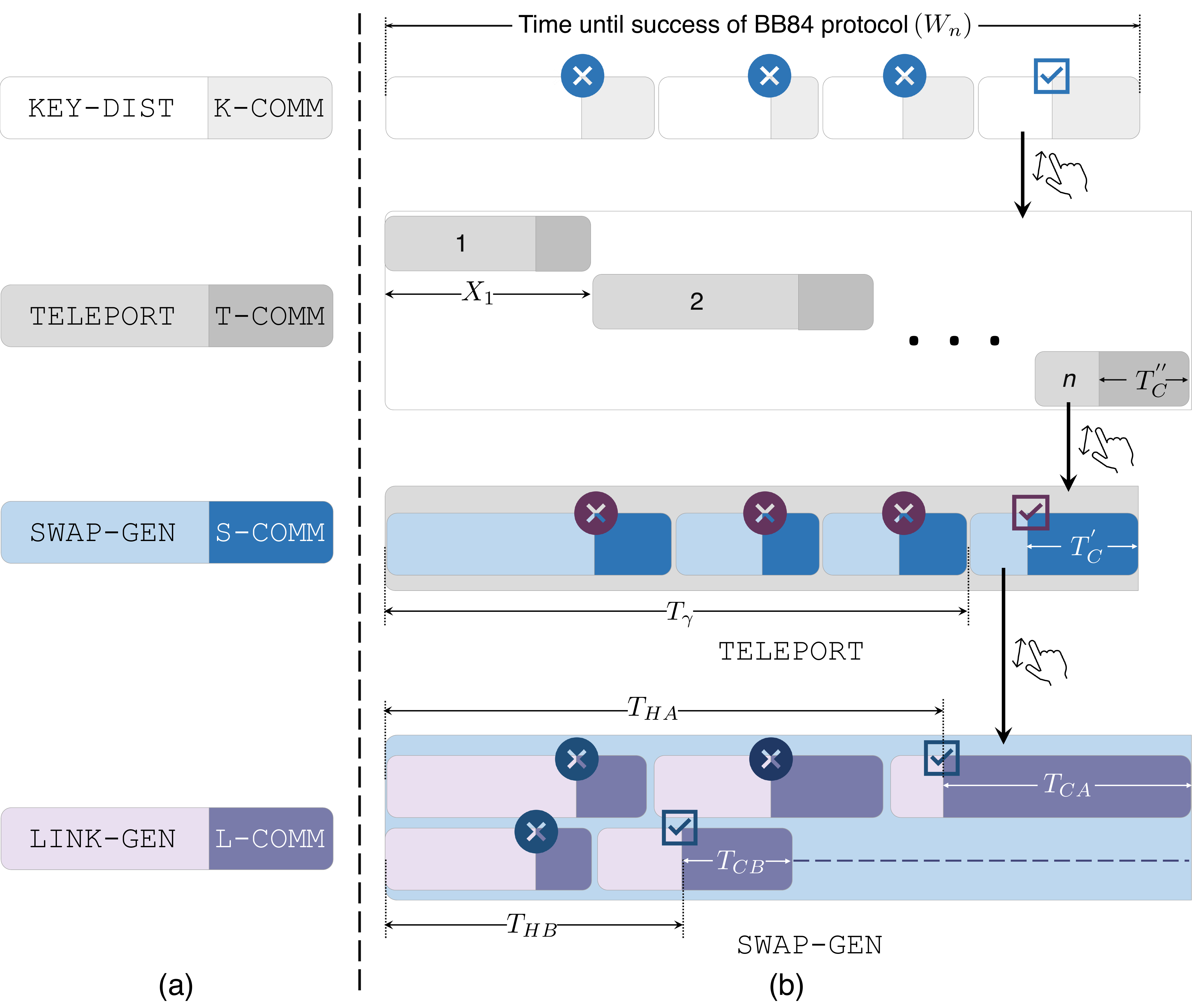}
    \caption{Illustration of the completion time of the BB84 protocol ($W_n$): (\textbf{a}) describes a trial unit, independent trials are repeated until success as shown in (\textbf{b}) except for the second row. The figure is not to scale. A BB84 attempt comprises two phases: teleportation of qubits (\texttt{KEY-DIST}) and reconciliation (\texttt{K-COMM}). \texttt{KEY-DIST} comprises $n$ sequential \texttt{TELEPORT+T-COMM} trials, each having a duration distributed as $X$. The MGF of $W_n$ relates to $X$ via~\eqref{eq:D}. Each trial within the \texttt{TELEPORT} unit is made of \texttt{SWAP-GEN} and \texttt{S-COMM} phases.  \texttt{S-COMM} represents communication of success/failure of a BSM at the repeater to the end-systems. Further, the duration of \texttt{SWAP-GEN} is the maximum of the durations of two modules: each representing the completion time of link-level entanglement generation whose trial unit comprises  \texttt{LINK-GEN} (link-level entanglement generation between an end-system and the repeater) and \texttt{L-COMM} (communication of success/failure of \texttt{LINK-GEN}) phases. The dotted line on the last row signifies the time one link has to wait for the other to succeed. Variables used for annotation are defined in Tab.~\ref{tab:notation} 
    .\looseness = -1}
    \label{fig:bb84diagram}
\end{figure*}

The set of elementary notations required for stating some of the main results is given in Tab.~\ref{tab:notation}.
A schematic description of the completion time of the protocol is shown in Fig.~\ref{fig:bb84diagram} in terms of the phases mentioned above. \looseness= -1


\subsection{Main Results}\label{sec:bb84Result}

In this section, we state the main result about the completion time of the BB84 protocol under assumtions~\ref{A1:singleChannel}~-~\ref{A6:durationsSE}.

\begin{theorem}[Completion time of version V1]\label{prop:mgfV1}
 The MGF of the completion time ($W_n$) of the BB84 protocol is given by:
 \begin{align}\label{eq:mgfWn}
  M_{W_n}(t) =  \frac{M_{K_C}(t)D_n(t)}{1 \! - \! M_{K_C}(t)(M_X^n(t)-D_n(t))}~,  
\end{align}
where $M_X^n(t)\! = \!(M_X(t))^n$ and
\begin{align}\label{eq:mgfWqkdFinal}
\begin{aligned}
    D_n (t)
    \overset{\Delta}{=}& \! \sum_{k=1}^{\ceil{\alpha n}} \!  M^{(1)}(t;k,\ceil{\beta k}) M_X^{n-k}(t) \\
    & \quad \frac{1}{2^n}\underset{\frac{k-1}{\alpha}<j\le\frac{k}{\alpha}}{\sum} \! \binom{n}{j}.
\end{aligned}
\end{align}
For $j \le l$, $M^{(1)}(t;l,j)$ is defined as follows:
\begin{align}\label{eq:an0ann}
\begin{aligned}
 M^{(1)}(t;l,0) &\overset{\Delta}{=} (m_0(t)+m_1(t))^l~, \\
 M^{(1)}(t;l,l) &\overset{\Delta}{=} m_1^l(t)~, \quad l \in \mathbb{N}~. 
\end{aligned}
\end{align}
Further, for $l \ge 2$, $1 \le j\le l-2$:
\begin{align}\label{eq:mgfWqkdOneSoln}
    &M^{(1)} \! (t;l,j) \! \overset{\Delta}{=} \! \sum_{k = 0}^1 \!  \binom{l-k}{j}m_0^{l-k-j}(t)m_1^{j+k}(t) \nonumber \\
    & \quad + m_1(t) \sum_{k = j}^{l-2}  (m_0(t)+ m_1(t))^{l-1-k} \\
    & \qquad \binom{k}{j} m_0^{k-j}(t) m_1^j(t) ~, \nonumber  
\end{align}
and
$$ M^{(1)}(t;l,l-1) \overset{\Delta}{=} m_1^l(t) + \binom{l}{1} m_0(t) m_1^{l-1}(t)~.$$
Note that $m_0^l(t)$ (resp. $m_1^l(t)$) means $(m_0(t))^l$ (resp. $m_1^l(t)$).
Further,
\begin{align}\label{eq:reDefm0m1}
  m_1(t) &\overset{\Delta}{=} \!\expect{e^{tX}Y} \!= \!\expect{e^{tX}|Y \!=\! 1} \prob{Y \!=\! 1}, \nonumber \\
m_0(t) &\overset{\Delta}{=}\! \expect{e^{tX}(1\!-\!Y)} \\
&=\! \expect{e^{tX}|Y \!=\! 0}\prob{Y \!=\! 0}~. \nonumber 
\end{align}
The RHS of~\eqref{eq:reDefm0m1} can be expressed in terms of the durations of the atomic phases of teleportation as follows:
\begin{align}\label{eq:m1}
    & m_1(t) = \frac{1}{4}M_{T_\gamma}(t)  \bigg(2I(t,\infty) M_{T_C^{'}}(t) M_{T_C^{''}}(t) \nonumber \\
    & \quad +w_0^2 I(t,t_c) M_{T_C^{'}}(t-\frac{1}{t_c}) \bigg(\! \!M_{T_C^{''}}(t\!-\!\frac{1}{t_{da}}) \nonumber \\
    & \quad +\!M_{T_C^{''}}\big(t\!-\!\frac{1}{t_{de}}\!-\!\frac{1}{2t_{da}}\big)\! \bigg)\!\bigg), \quad \text{with}\\
    & M_{T_\gamma}(t) = \frac{p_{\text{swap}}}{1\!-\!(1\!-\!p_{\text{swap}})M_{T_C^{'}}(t)I(t,\infty)}~, \nonumber 
\end{align}
and
\begin{align}\label{eq:I}
     I(t,s)\! \overset{\Delta}{=}\! \mathsf{E}\bigg(\!e^{t\max\{V_A,V_B\}} 
     e^{-\frac{|V_A-V_B|+T_{CA}+T_{CB}}{s}}\!\bigg). 
\end{align}
Also,
\begin{align}\label{eq:m0Mx}
\begin{aligned}
M_{X}(t) &= M_{T_\gamma}(t) I(t,\infty)M_{T_C^{'}}(t)M_{T_C^{''}}(t)~.  
\end{aligned} 
\end{align}
Since $m_0(t) = M_{X}(t)-m_1(t)$, we can compute it by plugging in expressions for $m_1(t)$ and $M_X(t)$.
The computation of $I(t,s)$ is described in Sect.~\ref{sec:numericalComp}.
\end{theorem}

Using Thm.~\ref{prop:mgfV1}, we can derive the CDF and a bound for the tail probability of $W_n$ as follows.

\begin{corollary}[CDF of $W_n$]\label{corr:CDF}
For $s>0$,
\begin{align}\label{eq:CDF}
\begin{aligned}
    \mathsf{P}(W_n \! \le \! s) &= \!  \mathcal{L}^{-1}\{F(t)\}(s), \thickspace \text{with} \\
    F(t) &\overset{\Delta}{=} \frac{M_{W_n}(-t)}{t} 
\end{aligned}
\end{align}
where $\mathcal{L}^{-1}$ denotes the inverse Laplace transform.
The numerical computation of~\eqref{eq:CDF} is described in Sect.~\ref{sec:numericalComp}.
\end{corollary}
\begin{proof}
    Let $f_n$ and $F_n$ be the density and CDF of $W_n$, respectively.
    Denoting Laplace transform of a function $h(t)$ by $\mathcal{L}\{h(t)\}(s)$, we have $\mathcal{L}\{f_n(t)\}(s) = M_{W_n}(-s)$.
    Since $F_n^{'} = f_n$ and $F_n(0) = 0$ (as $W_n>0$), $\mathcal{L}\{F_n(t)\}(s) = M_{W_n}(-s)/s$, which gives~\eqref{eq:CDF}.
\end{proof}

\begin{corollary}[Tail bound of the completion time of version V1]\label{prop:chfBrrWn}
Using Chernoff's method,
\begin{align}\label{eq:chernoffWn}
\mathsf{P}(W_n \! > \! s) \le \!  \inf_{t \in [0,b)} e^{-ts}M_{W_n}(t)~,
\end{align}    
where $b \!=\! \sup \{t \in \mathbb{R}: M_{W_n}(t)\!<\!\infty\}$.
Note that for any $t \in [0,b)$, $e^{-ts}M_{W_n}(t)$ is a valid upper bound for $\mathsf{P}(W_n \! > \! s)$ which we use for numerically computing the bound in Sect.~\ref{sec:numericalComp}.
\begin{proof}
This follows directly by applying Chernoff's method to $W_n$.
\end{proof}
\end{corollary}

%% file: sections/methodsNproofs.tex
\vspace{-10pt}
\section{Methods and Proofs of Results}\label{sec:methods}
\subsection{Background}\label{sec:rus}
\begin{figure*}[h!]
    \centering
    \includegraphics[width=\linewidth]{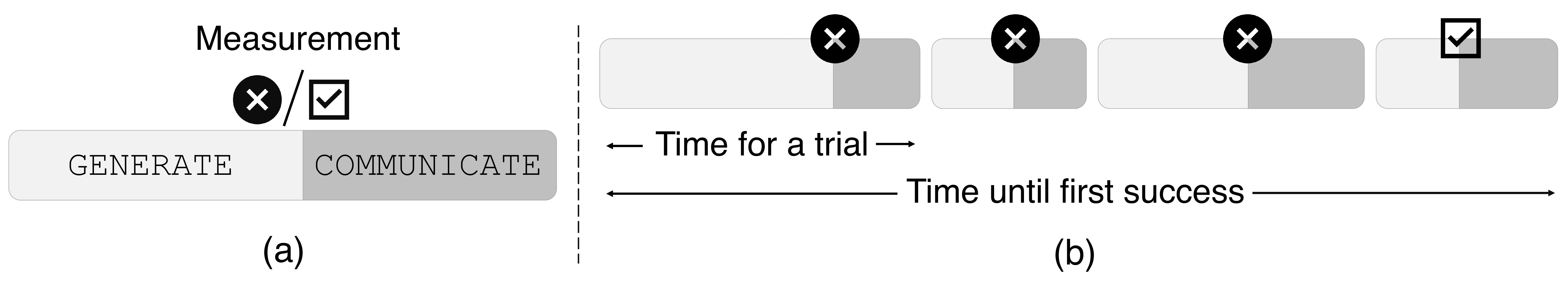}
    \caption{Schematic description of the completion time of a failure-prone protocol: (\textbf{a}) each trial begins with a \texttt{GENERATE} phase, followed by a measurement/detection event resulting in success (\checkmark) or failure ($\times$). The measurement outcome is in general dependent on granular details of the \texttt{GENERATE} phase and is communicated to relevant parties in the \texttt{COMMUNICATE}/\texttt{LOCC} phase. The durations of the two phases are random and independent. Independent trials are repeated until the measurement event results in success and the completion time is shown in (\textbf{b}).}
    \label{fig:RUS}
\end{figure*}

Following~\cite{brand, liRus, coopmans2022improved}, we first express the completion time $W$ of a failure-prone protocol in terms of the durations of its trial units.
We assume that each trial has a \texttt{GENERATE} phase, which depends on $r$ independent tasks.
We denote the duration of these tasks by $T_1, T_2, \dots, T_r$.
Once the tasks finish, there is an instantaneous measurement/detection event which succeeds with probability $p(L_1, L_2, \dots, L_r,\theta)$, where $\{L_i\}_{i \in [r]}$ are latent variables that completely determine $\{T_i\}_{i \in [r]}$ and $\theta$ is a parameter reflecting the efficiency of the underlying physical system.
If the measurement event results in failure, the process starts over immediately and carries on until success is observed. 
Note that the success or failure of the event has to be duly communicated so that the process terminates or moves on to the next trial.
We call this part of a trial the \texttt{COMMUNICATE} phase and denote its duration by $T_C$.
We assume that $T_C$ is independent of $\{L_i\}_{i \in [r]}$ and that the trials are independent as well.
In case the measurement event takes non-negligible time, we can consider $T_C$ to be the total time corresponding to the measurement and communication events.
In this case, the  \texttt{COMMUNICATE} phase is termed as \texttt{LOCC} phase.
For a schematic description of the completion time, see Fig.~\ref{fig:RUS}.

\begin{figure}
    \centering
\makebox[0.95\columnwidth]{\fbox{
\begin{minipage}{0.95\columnwidth}
  {\vspace{1pt}
  \small{
  \noindent $W$: the completion time of the protocol, \newline
  $T_1, T_2, \dots, T_r$: the durations of the constituent tasks of the \texttt{GENERATE} phase,\newline
  $L_1, L_2, \dots, L_r$: the latent variable completely determining the duration of the \texttt{GENERATE} phase and success probability of the measurement event, \newline
  $\theta$: hardware efficiency parameter for measurement,\newline
  $p(L_1, L_2, \dots, L_r,\theta)$: the success probability of the measurement event,\newline
  $T_C$: \texttt{COMMUNICATE}/\texttt{LOCC} phase duration.}}
\captionof{table}{Notations used to describe a failure-prone protocol in Sect.\ref{sec:rus}.}
\label{tab:notation3}
\end{minipage}
}}
\end{figure}

\begin{prop}\label{prop:mgfW}
(i) Let us assume that the $r$ tasks in the \texttt{GENERATE} phase run in parallel. The MGF of the completion time is then given by: \looseness = -1
\begin{align}\label{eq:mgfW}
\begin{aligned}
    &M_W(t) \\
    = & \frac{\mathsf{E}(e^{t T_C}) \mathsf{E}\big(e^{t \max_{i \in [r]}T_i} p_{\theta}(\mathbf{L}^{(1)})\big)}{1-\mathsf{E}(e^{t T_C}) \mathsf{E}\big(e^{t \max_{i \in [r]}T_i} (1-p_{\theta}(\mathbf{L}^{(1)}))\big)}~,
\end{aligned}
\end{align}
for $\mathsf{E}(e^{t T_C}) \mathsf{E}\big(e^{t \max_{i \in [r]}T_i} (1-p_{\theta}(\mathbf{L}^{(1)}))\big)<1$.
Here, $[r] = \{1,2,\dots,r\}$, $\mathbf{L}^{(k)}=(L_1^{(k)}, L_2^{(k)}, \dots, L_r^{(k)})$ is the vector comprising the latent variables of the $r$ constituent tasks of the \texttt{GENERATE} phase for the $k$-th trial, and 
$$\mathbf{L}^{(k)} \overset{iid}{\sim}(L_1, L_2, \dots, L_r)~.$$

(ii) If the $r$ tasks run in sequence instead, 
\begin{align}\label{eq:mgfWSeq}
    M_W(t)
    \!=\! \frac{M_{T_C}(t) \mathsf{E}\big(e^{t \sum_{i = 1}^r\! T_i} p_{\theta}(\mathbf{L}^{(1)})\big)}{1\!-\!M_{T_C}(t) \mathsf{E}\!\big(e^{t \sum_{i = 1}^r\! T_i} (1\!-\!p_{\theta}(\mathbf{L}^{(1)}))\big)}.
\end{align}
\end{prop}

\begin{proof}
(i) If the tasks run in parallel, the duration of each trial is given by: $U =\max_{i \in [r]}T_i+T_C$.
Further, if we denote the success of the measurement event as $Y$, 
$$Y|\{L_i\}_{i \in [r]} \sim \text{Bern}(p(\{L_i\}_{i \in [r]},\theta)).$$
Let $N$ denote the number of trials until success.
Using the shorthand $p(\mathbf{L},\theta) = p_{\theta}(\mathbf{L})$,
\begin{align}\label{eq:condMgfW}
\begin{aligned}
    &\mathsf{E}\big(e^{tW} \mathbbm{1}_{N = k}|\{\mathbf{L}^{(j)}\}_{j=1}^\infty\big)\\
    = & e^{t\sum_{j = 1}^k U_j} p_{\theta}(\mathbf{L}^{(k)}) \prod_{j=1}^{k-1}\big(1-p_{\theta}(\mathbf{L}^{(j)})\big),
\end{aligned}
\end{align}
where $U_j = \max_{i \in [r]}\{T_i^{(j)}\}+T_C^{(j)}.$ 
Since the RHS of~\eqref{eq:condMgfW} is non-negative, summing over $k$ and taking expectation with respect to $\{\mathbf{L}^{(j)}\}_{j=1}^\infty$, we have:
\begin{align*}
    &M_W(t) \\
    =& \sum_{k=1}^\infty\! \mathsf{E}\big(e^{t U_1} p_{\theta}(\mathbf{L}^{(1)})\big) \mathsf{E}^{k-1}\! \big(e^{t U_1}\! (1\!-\!p_{\theta}(\mathbf{L}^{(1)}))\big) \\
    =& \frac{\mathsf{E}\big(e^{t U_1} p_{\theta}(\mathbf{L}^{(1)})\big)}{1-\mathsf{E}\big(e^{t U_1} (1-p_{\theta}(\mathbf{L}^{(1)}))\big)} \\
    =& \frac{\mathsf{E}(e^{t T_C}) \mathsf{E}\big(e^{t \max_{i \in [r]}T_i} p_{\theta}(\mathbf{L}^{(1)})\big)}{1-\mathsf{E}(e^{t T_C}) \mathsf{E}\big(e^{t \max_{i \in [r]}T_i} (1-p_{\theta}(\mathbf{L}^{(1)}))\big)}~, 
\end{align*}
for $\mathsf{E}(e^{t U_1} (1-p_{\theta}(\mathbf{L}^{(1)})))<1$, as claimed.

(ii) The result follows simply by observing that $U_j = \sum_{i = 1}^r T_i^{(j)}+T_C^{(j)}$ when tasks are executed sequence and imitating the proof of part (i).
\end{proof}

The next results are useful for deriving tail bounds for the completion time $W$.
\begin{prop}\label{prop:existenceMgf}
Let $I_1, I_2,\dots, I_r, I_C$ be the neighbourhoods of zero where the MGFs of the RVs $T_1, T_2,\dots, T_r, T_C$ exist, respectively. 
Further, let $I_0$ denote the neighbourhood of zero where $G(t) = \mathsf{E}(e^{t U_1} (1-p_{\theta}(\mathbf{L}^{(1)})))<1$.
Then the MGF of $W$ exists in $\bigcap_{l \in [r]\cup\{0,C\}} I_l$.
\end{prop}
\begin{proof}
    See Sect.~\ref{sec:defProofs}.
\end{proof}

\begin{corollary}\label{prop:subExp}
If $T_1,T_2,\dots,T_r,T_C$ are sub-exponential, so is $W$.
\end{corollary}
\begin{proof}
    See Sect.~\ref{sec:defProofs}.
\end{proof}

Corollary~\ref{prop:subExp} essentially guarantees the existence of the MGF of $W$ in a neighbourhood of zero, which implies that we can apply Chernoff's method to obtain an upper bound for the tail probability:
\begin{align}\label{eq:chernoffBound}
    \mathsf{P}(W \! > \! s) \!  &\le \!  \inf_{t \in [0,b)} e^{-ts}\mathsf{E}(e^{tW})~,
\end{align}
where $b = \sup \{t \in \mathbb{R}: M_{W}(t)< \infty\}.$

\subsection{Proofs of Main Results}\label{sec:proofsMain}
We organise this section as follows: we first derive the success probability of a teleportation attempt given the Werner parameter of the entanglement resource and the time it takes for Alice to send her measurement result to Bob during teleportation.
The result is subsequently used to show that the expressions of $m_1$ and $M_X$ in~\eqref{eq:m1} and~\eqref{eq:m0Mx} are valid.
Next, we derive the MGF of the completion time of the simplified version called V0, which helps us compute the MGF for the actual version as stated in Thm.~\ref{prop:mgfV1}. \looseness = -1

\begin{widetext}
\begin{lemma}\label{lemma:successProb}
    Let us assume that teleportation starts with an entanglement resource given by the Werner parameter $w$. We also denote the time required for Alice to send the measurement result to Bob during teleportation by $T_C^{''}$ (see Tab.~\ref{tab:notation}). The success probability of teleportation given these quantities can then be expressed as:
    \begin{align*}
    p_\Lambda \overset{\Delta}{=} &\prob{Y=1|w,T_C^{''}} \\
    = &\frac{1}{4}\big(2+we^{-T_C^{''}/t_{da}} + we^{-T_C^{''}(1/t_{de}+1/2t_{da})}\big)~.
    \end{align*}
Recall that $t_{de}$ and $t_{da}$ respectively denotes the constants characterising the effect of the dephasing and the amplitude damping noise on Bob's quantum memory.
\end{lemma}
\begin{proof}
We start with the density matrix representation of the joint state of Alice and Bob through the stages of standard teleportation protocol of the data qubit $\ket{\phi} = a \ket{0}+b \ket{1}$, which may be found in standard textbooks, especially considering $\ket{\Phi^+} = \frac{1}{\sqrt{2}}(\ket{00}+\ket{11})$ as the entanglement resource.
Below we consider that Alice holds the data qubit $\ket{\phi}$ and $\ket{\Psi^+} = \frac{1}{\sqrt{2}}(\ket{01}+\ket{10})$ is shared between Alice and Bob.
This is a necessary step towards the calculation of the success probability of a teleportation attempt because, in the BB84 key distribution example, teleportation starts with an entanglement resource given by a general Werner state instead of $\ket{\Phi^+}$.
As per convention, the first and second qubits belong to Alice, whereas the third belongs to Bob.
If not written in the natural order, the qubit indices are written explicitly outside a parenthesis containing the relevant state.
\normalsize{The stages of standard teleportation protocol with $\ket{\Psi^+}$ can be written as follows:}
\small{
\begin{align}\label{eq:stagesOfTp}
\begin{aligned} 
    & \frac{1}{2}\big(\ket{01}\!\bra{01}\!+\!\ket{01}\!\bra{10}\!+\!\ket{10}\!\bra{01}\!+\!\ket{10}\!\bra{10}\big)_{13}\big(a^2\!\ket{0}\!\bra{0}\!+\!a^*b\!\ket{1}\!\bra{0}\!+\!ab^*\!\ket{0}\!\bra{1}\!+\!b^2\!\ket{1}\!\bra{1}\big)_2 \\
    = & \frac{a^2}{2}\big(\ket{001}\!\bra{001}\!+\!\ket{001}\!\bra{100}\!+\!\ket{100}\!\bra{001}\!+\!\ket{100}\!\bra{100}\big)
    \!+\!\frac{a^*b}{2}\big(\ket{011}\!\bra{001}\!+\!\ket{011}\!\bra{100}\!+\!\ket{110}\!\bra{001}\\
     \thickspace & 
    +\!\ket{110}\!\bra{100}\big)
    \!+\!\frac{ab^*}{2}\big(\ket{001}\!\bra{011}\!+\!\ket{001}\!\bra{110}\!+\!\ket{100}\!\bra{011}\!+\!\ket{100}\!\bra{110}\big)
    \!+\!\frac{b^2}{2}\big(\ket{011}\!\bra{011}\!+\!\ket{011}\!\bra{110}\\
     \thickspace & +\!\ket{110}\!\bra{011}\!+\!\ket{110}\!\bra{110}\big) \\
     \xrightarrow{O_1} & \frac{a^2}{2}\big(\ket{001}\!\bra{001}\!+\!\ket{001}\!\bra{110}\!+\!\ket{110}\!\bra{001}\!+\!\ket{110}\!\bra{110}\big)\!+\!\frac{a^*b}{2}\big(\ket{011}\!\bra{001}\!+\!\ket{011}\!\bra{110}\!+\!\ket{100}\!\bra{001} \\
     \thickspace & +\!\ket{100}\!\bra{110}\big)\!+\!\frac{ab^*}{2}\big(\ket{001}\!\bra{011}\!+\!\ket{001}\!\bra{100}\!+\!\ket{110}\!\bra{011} \!+\! \ket{110}\!\bra{100}\big) \!+\! \frac{b^2}{2}\big(\ket{011}\!\bra{011}\!+\!\ket{011}\!\bra{100}\\
     \thickspace & +\!\ket{100}\!\bra{011}\!+\!\ket{100}\!\bra{100}\big) \\
     \xrightarrow{O_2} & \frac{a^2}{4}\big(\!\ket{001}\!\bra{001}\!+\!\ket{001}\!\bra{110}\!+\!\ket{110}\!\bra{001}\!+\!\ket{110}\!\bra{110}\!+\! \ket{011}\!\bra{001}\!+\!\ket{011}\!\bra{110}\!+\!\ket{100}\!\bra{001}\!+\!\ket{100}\!\bra{110}\\     \thickspace & +\ket{001}\!\bra{011}\!+\!\ket{001}\!\bra{100}\!+\!\ket{110}\!\bra{011}\!+\!\ket{110}\!\bra{100}\!+\! \ket{011}\!\bra{011}\!+\!\ket{011}\!\bra{100}\!+\!\ket{100}\!\bra{011}\!+\!\ket{100}\!\bra{100}\big) \\     \thickspace & 
     +\frac{a^*b}{4}\big(\dots\big)+\frac{ab^*}{4}\big(\dots\big)+\frac{b^2}{4}\big(\dots\big)\\     
     = & \frac{1}{4}\ket{00}\!\bra{00}\big(a^2\!\ket{1}\!\bra{1}\!+\!a^*b\!\ket{0}\!\bra{1}\!+\!ab^*\!\ket{1}\!\bra{0}\!+\!b^2\!\ket{0}\!\bra{0}\big)\!+\!
     \frac{1}{4}\ket{01}\!\bra{01}\big(a^2\!\ket{0}\!\bra{0}\!+\!a^*b\!\ket{1}\!\bra{0} \!+\!ab^*\!\ket{0}\!\bra{1} \\     \thickspace &+\!b^2\!\ket{1}\!\bra{1}\big) +
     \frac{1}{4}\ket{10}\!\bra{10}\big(a^2\!\ket{1}\!\bra{1}\!-\!a^*b\!\ket{0}\!\bra{1}\!-\!ab^*\!\ket{1}\!\bra{0}\!+\!b^2\!\ket{0}\!\bra{0}\big)\!+\!
     \frac{1}{4}\ket{11}\!\bra{11}\big(a^2\!\ket{0}\!\bra{0} \!-\!a^*b\!\ket{1}\!\bra{0} \\ 
     \thickspace &-\!ab^*\!\ket{0}\!\bra{1}\!+\!b^2\!\ket{1}\!\bra{1}\big)+\rho_0~,
\end{aligned}
\end{align}
}
\normalsize{where $\rho_0$ denotes the residual terms that do not make any contribution to the final state when measured in the standard basis.}
Further, the operations $O_1$ and $O_2$ denote $\text{CNOT}_{12}$ and $H_1 \otimes I_2$, respectively.
Therefore, if teleportation starts with $p_1\ket{\Phi^+}\!\bra{\Phi^+}\!+\!p_2\ket{\Phi^-}\!\bra{\Phi^-}\!+\!p_3\ket{\Psi^+}\!\bra{\Psi^+}\!+\!p_4\ket{\Psi^-}\!\bra{\Psi^-}$ as the entanglement resource, the (unnormalised) state of the system immediately before Alice's measurement is given by:
\begin{align}\label{eq:stateB4measure}
\begin{aligned}
    & \ket{00}\!\bra{00}\big(p_1\!\ket{\phi}\!\bra{\phi} \!+\! p_2\!\ket{\phi_z}\!\bra{\phi_z} \!+\! p_3\!\ket{\phi_x}\!\bra{\phi_x}\!+\!p_4\!\ket{\phi_y}\!\bra{\phi_y}\big)\!+\!    \ket{01}\!\bra{01}\big(p_1\!\ket{\phi_x}\!\bra{\phi_x} \!+\! p_2\!\ket{\phi_y}\!\bra{\phi_y} \\
    & \thickspace +\! p_3\!\ket{\phi}\!\bra{\phi} \!+\! 
    p_4\!\ket{\phi_z}\!\bra{\phi_z}\big)\!+\! \ket{10}\!\bra{10}\big(p_1\!\ket{\phi_z}\!\bra{\phi_z} \!+\! p_2\!\ket{\phi}\!\bra{\phi} \!+\! p_3\!\ket{\phi_y}\!\bra{\phi_y}\!+\!p_4\!\ket{\phi_x}\!\bra{\phi_x}\big) \\
    & \thickspace
    +\! \ket{11}\!\bra{11}\big(p_1\!\ket{\phi_y}\!\bra{\phi_y} \!+\! p_2\!\ket{\phi_x}\!\bra{\phi_x} \!+\! p_3\!\ket{\phi_z}\!\bra{\phi_z}\!+\!p_4\!\ket{\phi}\!\bra{\phi}\big)+\rho_{00}~,
\end{aligned}
\end{align}
where $\rho_{00}$ denotes the terms not contributing to measurement results and $\ket{\phi_x},\ket{\phi_y}$, and $\ket{\phi_z}$ denote Pauli X, XZ, and Z rotations of $\ket{\phi}$, respectively.
Let us now denote the state of Bob's qubit immediately after Alice's measurement as $\rho(\ket{\phi})$.
Since the communication of the measurement outcome takes time $T_C^{''}$, the state of Bob's qubit before applying a unitary operation ($Z^l X^k$ for some $l,k \in \{0,1\}$ according to the outcome) is given by $\mathcal{N}_{T_C^{''}}(\rho)$. \looseness = -1

Recall that in the BB84 protocol, the data qubit $\ket{\phi} \in \{\ket{0},\ket{1},\ket{+},\ket{-}\}$ and for an entanglement resource specified by Werner parameter $w$, $p_1 = (1+3w)/4$, $p_2 = p_3 = p_4 = (1-w)/4$.
Substituting the values of $\ket{\phi}$ and $\{p_i\}_{i=1}^4$ in~\eqref{eq:stateB4measure}, 
the state $\rho(\ket{\phi})$ can be further simplified to:
$$\rho(\ket{\phi}) = \frac{1+w}{2}\ket{\phi}\bra{\phi}+\frac{1-w}{2}\ket{\phi'}\bra{\phi'} \quad \text{or} \quad \frac{1-w}{2}\ket{\phi}\bra{\phi}+\frac{1+w}{2}\ket{\phi'}\bra{\phi'}~,$$
each with probability $1/2$.
Here $\ket{\phi'}$ maps $(\ket{0},\ket{1},\ket{+},\ket{-})$ to $(\ket{1},\ket{0},\ket{-},\ket{+})$.
Note that Bob applies a suitable unitary transformation ($Z^l X^k$) on $\mathcal{N}_{T_C^{''}}(\rho(\ket{\phi}))$ to recover $\ket{\phi}\bra{\phi}$.
Further,
\begin{align*}
    \mathcal{N}_{T_C^{''}}(\ket{0}\!\bra{0}) \!&= \!\ket{0}\!\bra{0}~, \quad
    \mathcal{N}_{T_C^{''}}(\ket{1}\!\bra{1}) \!=\! \gamma(\cdot)\ket{0}\!\bra{0}+\big(1-\gamma(\cdot)\big)\ket{1}\!\bra{1}~, \\
    \mathcal{N}_{T_C^{''}}(\ket{\phi}\!\bra{\phi}) \!&=\! \frac{1\!+\!(1\!-\!2p(\cdot))\sqrt{1\!-\!\gamma(\cdot)}}{2}\ket{\phi}\!\bra{\phi}\!+\!\frac{1\!-\!(1\!-\!2p(\cdot))\sqrt{1\!-\!\gamma(\cdot)}}{2}\ket{\phi'}\!\bra{\phi'}, \ket{\phi}\! \in \! \{\ket{+},\ket{-}\},
\end{align*}
where we have omitted the argument $T_C^{''}$ in $\gamma(T_C^{''})$ and $p(T_C^{''})$ for brevity.
This implies:
\begin{align*}
    \mathcal{N}_{T_C^{''}}(\rho(\ket{0})) =
    \begin{cases}
        \bigg(\frac{1+w}{2}+\gamma(\cdot)\frac{1-w}{2}\bigg)\ket{0}\bra{0}+(1-\gamma(\cdot))\frac{1-w}{2}\ket{1}\bra{1}, & \thickspace \text{w.p.} ~ \frac{1}{2} \\
        \bigg(\frac{1-w}{2}+\gamma(\cdot)\frac{1+w}{2}\bigg)\ket{0}\bra{0}+(1-\gamma(\cdot))\frac{1+w}{2}\ket{1}\bra{1}, & \thickspace \text{w.p.} ~ \frac{1}{2}.
    \end{cases}   
\end{align*}
We can now calculate the probability of Bob recovering $\ket{0}$ as $(1+w-w \gamma(\cdot))/2$, which turns out to be the recovery probability for $\ket{\phi} = \ket{1}$ as well.
On the other hand, Bob successfully recovers $\ket{+}$ or $\ket{-}$ with probability $(1\!+\!w(1\!-\!2p(\cdot))\sqrt{1\!-\!\gamma(\cdot)})/2$.
Since all four data qubits are equally likely in the BB84 protocol, we have
\begin{align*}
    p_\Lambda = &\frac{1}{4}\big( 1+w-w\gamma(T_C^{''})+1+ w (1-2p(T_C^{''}))\sqrt{1-\gamma(T_C^{''})}\big)  \\
    = & \frac{1}{4}\big(2+we^{-T_C^{''}/t_{da}} + we^{-T_C^{''}(1/t_{de}+1/2t_{da})}\big)~.
\end{align*}
\end{proof}
\vspace{-5pt}
\end{widetext}

\begin{proof}[Proof of~\eqref{eq:m1} and~\eqref{eq:m0Mx}] 
First, we refer the reader to Fig.~\ref{fig:bb84diagram}, where the teleportation duration of a single qubit is shown and its constituent phases and respective durations are annotated.
Recall that $T_{HA}$ (resp. $T_{HB}$) denotes the time until Alice (resp. Bob) successfully generates a link-level entanglement and the communication of this event takes $T_{CA}$ (resp. $T_{CB}$) amount of time.
To derive the expression for $M_{T_{\gamma}}(t)$ in~\eqref{eq:m1}, we first see that the swap failure time $T_{\gamma}$ comprises $(\text{Geo}(p_{\text{swap}})-1)$ independent swap trials, each having a duration of $\max\{T_{HA}+T_{CA},T_{HB}+T_{CB}\}+T_C^{'}$.
Applying~\eqref{eq:mgfW} now leads to the given expression.

The expression for $M_X(t)$ in~\eqref{eq:m0Mx} now follows from the break-down of the qubit teleportation time $X$ given in Tab.~\ref{tab:notation}:
\begin{align} \label{eq:teleportTime}
  X = T_\gamma+\max\{V_A,V_B\}+T_C^{'}+T_C^{''}~,
\end{align}
 the definition of $I(t,s)$ in~\eqref{eq:I}, and the fact that the components in the RHS above are independent.

Next, we establish the formula for $m_1(t)$. 
Due to assumption~\ref{A4:linkModelling}, a successful swap thus leads to an end-to-end entanglement between Alice and Bob, characterised by the Werner parameter:
\begin{align*}
  w_0^2 e^{-(|T_{HA}+T_{CA}\!-\!T_{HB}\!-\!T_{CB}|+T_{CA}+T_{CB})/t_c}~,
\end{align*}
where $w_0$ denotes the Werner parameter of a freshly generated link and $t_c$ denotes the joint coherence time of the involved memories. 
Teleportation takes further $T_C^{'}$ time to start and thus begins with the following entanglement resource:
\begin{align*}
  w = w_0^2 e^{-(|V_A-V_B|+T_{CA}+T_{CB}+T_C^{'})/t_c}~.
\end{align*}
Recall from Tab.~\ref{tab:notation} that $V_A = T_{HA}+T_{CA}$ (resp. for B).
The decoherence of Bob's qubit while Alice sends the measurement results (during the time span $T_C^{''}$) is captured by the noise model described in assumption~\ref{A5:noiseModel}.
According to Lemma~\ref{lemma:successProb}, given $(T_{HA},T_{CA},T_{HB},T_{CB},T_C^{'},T_C^{''})$, the teleportation attempt succeeds with probability 
\begin{align} \label{eq:pLambda}
        p_\Lambda = \frac{1}{4}\big(2\!+\!we^{-T_C^{''}/t_{da}} \!+\! we^{-T_C^{''}(\frac{1}{t_{de}}+\frac{1}{2t_{da}})}\big).
\end{align}

Since $m_1(t) = \expect{e^{tX}Y}$, we can now use the expression for $X$ and $\prob{Y=1|w,T_C^{''}}$ to calculate the expectation, which leads to~\eqref{eq:m1}.
\end{proof}


We now prove Lemma~\ref{lemma:mgfV0}, which helps us express the MGF of the completion time of the actual version V1 as formalised in Thm.~\ref{prop:mgfV1}.

\begin{lemma}[Completion time of version V0]\label{lemma:mgfV0}
We define $c = \ceil{\beta n}$.
The MGF of the completion time ($W_{n,c}$) of version V0 is then given by:
$$M_{W_{n,c}}(t) = \frac{M_{K_C}(t)M^{(1)}(t;n,c)}{1 \! - \! M_{K_C}(t)M^{(0)}(t;n,c)}~,$$
where $M^{(1)}$ is defined in Thm.~\ref{prop:mgfV1} and
\begin{align}\label{eq:mgfWqkdZeroSoln}
\begin{aligned}
    M^{(0)}(t;l,j) = (m_0(t)+m_1(t))^l - M^{(1)}(t;l,j).
\end{aligned}
\end{align}
\end{lemma}

\begin{proof}
Recall that in version V0, Bob measures all teleported qubits in the same bases as Alice encoded them and all of them are used for reconciliation.
Let us denote by $X_i$ the duration of the teleportation of the $i$-th qubit. Further, let $Y_i$ be the indicator variable assuming the value $1$ when Bob's measurement result for the $i$-th qubit agrees with what Alice originally encoded.
Therefore, $(X_i,Y_i)\overset{iid}{\sim}(X,Y)$.
Further, the duration of a BB84 attempt is given by: $K_C+\sum_{i=1}^n X_i$.
Now, let $c$ be the success threshold of the protocol, i.e., a BB84 attempt succeeds iff $\sum_{i=1}^n Y_i \ge c$.
Thus, using part (ii) of Prop.~\ref{prop:mgfW}, we have:
\begin{align}\label{eq:mgfWqkd}
    M_{W_{n,c}}(t)
      \! &= \! \frac{M_{K_C}\!(t)\mathsf{E}(e^{t \sum_{i=1}^n X_i}\mathbbm{1}_{\sum_{i=1}^n Y_i \ge c})}{1 \! - \! M_{K_C}\!(t)\mathsf{E}(e^{t \! \sum_{i=1}^n \! X_i}\mathbbm{1}_{\sum_{i=1}^n \!Y_i < c})}.
\end{align}
Let us define 
\begin{align*}
    \bar{M}^{(1)}(t;l,j) &\overset{\Delta}{=} \mathsf{E}(e^{t \sum_{i=1}^l X_i}\mathbbm{1}_{\sum_{i=1}^l Y_i \ge j})~,\\
    \bar{M}^{(0)}(t;l,j) &\overset{\Delta}{=} \mathsf{E}(e^{t \! \sum_{i=1}^l \! X_i}\mathbbm{1}_{\sum_{i=1}^l \!Y_i < j})~.
\end{align*}
To complete the proof, we need to show that $\bar{M}^{(1)}\!\equiv\!{M}^{(1)}$ and $\bar{M}^{(0)}\!\equiv\!{M}^{(0)}$.
Note that $\bar{M}^{(1)}(t;1,0)\!=\!m_0(t)\!+\!m_1(t)$ and $\bar{M}^{(1)}(t;1,1)\!=\!m_1(t)$.
Since $(X_i,Y_i)\!\overset{iid}{\sim}\!(X,Y)$,
\begin{align}\label{eq:an0annNew}
\begin{aligned}
 \bar{M}^{(1)}(t;l,0) &= (m_0(t)+m_1(t))^l~, \\
 \bar{M}^{(1)}(t;l,l) &= m_1^l(t)~, \quad l \in \mathbb{N}~.    
\end{aligned}
\end{align}
Now, for $l \ge 2$, $1 \le j\le l$,
\begin{align}\label{eq:mgfWqkdOne}
    & \bar{M}^{(1)}(t;l,j)\nonumber \\
    & = \mathsf{E}\big(e^{t \sum_{i=1}^l X_i}(
    \mathbbm{1}_{Y_1 = 0}\mathbbm{1}_{\sum_{i=2}^l Y_i \ge j}\nonumber \\
    & \quad +\mathbbm{1}_{Y_1 = 1}\mathbbm{1}_{\sum_{i=2}^l Y_i \ge j-1})\big)\\
    & = \mathsf{E}(e^{t X_1}\mathbbm{1}_{Y_1 = 0}) \mathsf{E}(e^{t \sum_{i=2}^l X_i} \mathbbm{1}_{\sum_{i=2}^l Y_i \ge j}) \nonumber \\
    & \quad + \!  \mathsf{E}(e^{t X_1} \! \mathbbm{1}_{Y_1 = 1}) \mathsf{E}(e^{t\! \sum_{i=2}^l \! X_i} \! \mathbbm{1}_{\sum_{i=2}^l  \! Y_i \ge j-1}\! ) \nonumber \\
    & = m_0(t) \bar{M}^{(1)}(t;l-1,j) \nonumber \\
    & \quad +m_1(t) \bar{M}^{(1)}(t;l-1,j-1)~. \nonumber
\end{align}
Here, the second and the third equality follow from the fact that $(X_i,Y_i)\!\overset{iid}{\sim}\!(X,Y)$. 

To solve the recurrence relation, let us introduce the following shorthand: $a_{l,j}=\bar{M}^{(1)}(t;l,j)$, $m_0 = m_0(t)$, and $m_1 = m_1(t)$
Therefore, $a_{1,0}=m_0+m_1$, $a_{1,1}=m_1$, and for $l \ge 2$, $1 \le j \le l-1$,
\begin{align}\label{eq:ancRecursion}
  a_{l,j} = m_0 a_{l-1,j} + m_1 a_{l-1,j-1}~.  
\end{align}
Also, \eqref{eq:an0annNew} can be rewritten as:
\begin{align}\label{eq:an0ann_}
 a_{l,0} \!=\! (m_0\!+\!m_1)^l, \quad
 a_{l,l} \!=\! m_1^l, \quad l \in \mathbb{N}.
\end{align}
We now define the generating function:
$$A_l(x) \overset{\Delta}{=} \sum_{j = 0}^l a_{l,j} x^j, \quad l \in \mathbb{N}~.$$
Multiplying both sides of~\eqref{eq:ancRecursion} by $x^j$ and summing over $1\le j\le l-1$, we have for $l \ge 2$:
\begin{align}\label{eq:AnRec}
\begin{aligned}
    A_l(x)= & (m_0+m_1 x) A_{l-1}(x) \\
            & \quad +(m_0+m_1)^{l-1} m_1~, 
\end{aligned}
\end{align}
with $A_1(x) = m_0+m_1+m_1 x$. 
Note that we have used~\eqref{eq:an0ann_} to arrive at~\eqref{eq:AnRec}.
For $l \ge 2$, recursive substitution in~\eqref{eq:AnRec} leads to:
\begin{align*}
    & A_l(x) = (m_0+m_1x)^{l-1} A_1(x) \\
    & \thickspace +m_1 \sum_{k = 0}^{l-2} (m_0+m_1 x)^k (m_0+m_1)^{l-1-k}~, \nonumber
\end{align*}
which, after some calculations, yields:
\begin{align*}
\begin{aligned}
    & a_{l,l-1} = m_1^l + \binom{l}{1} m_0 m_1^{l-1}~,\\
    & a_{l,j}\!=\!\binom{l}{j}m_0^{l-j}m_1^j\!+\! \binom{l-1}{j}m_0^{l-1-j}m_1^{j+1} \\
    & \thickspace + m_1 \sum_{k = j}^{l-2} (m_0+m_1)^{l-1-k} \binom{k}{j} m_0^{k-j} m_1^j~,
\end{aligned}
\end{align*}
for $1\le j\le l-2$.
It is now straightforward to see that $\bar{M}^{(1)}\equiv{M}^{(1)}$.
Further, $\expect{e^{tX}} = m_0(t)+m_1(t)$, which establishes $\bar{M}^{(0)}\equiv{M}^{(0)}$.
\end{proof}

\begin{proof}[Proof of Thm.~\ref{prop:mgfV1}] 
Recall that out of the $B \sim \text{Bin}(n,1/2)$ qubits where Bob's measurement bases agree with Alice's, $B_1 = \ceil{\alpha B}$ are sampled without replacement for checking the correctness, and the protocol is deemed successful if the corresponding measurements match for at least $\beta$ fraction of the sample.
The protocol has to be rerun until $B_1 \ge 1$.
Let $Z_i$ denote the Bernoulli RV assuming value $1$ iff the $i$-th qubit is sampled, i.e., $B_1 = \sum_{i=1}^n  Z_i$. 
Further, let $k_i$'s for $i \in [B_1]$ be the indices such that $Z_{k_i} = 1$.
Therefore, the protocol is successful when $B_1 \ge 1$ and $\sum_{i=1}^{B_1}\! Y_{k_i} \ge \beta B_1$, where $Y_j$ is the indicator variable that takes value $1$ when Bob's measurement for the $j$-th qubit agrees with its original value.
Denoting
\begin{align}\label{eq:D}
    \bar{D}_n(t) \!\overset{\Delta}{=}\! \mathsf{E}(e^{t \sum_{1}^n\! X_i}\mathbbm{1}_{\sum_{i=1}^{B_1}\! Y_{k_i} \ge \beta B_1}\mathbbm{1}_{B_1 \ge 1}),
\end{align}
we have:
\begin{align*}
  M_{W_n}(t) =  \frac{M_{K_C}(t)\bar{D}_n(t)}{1 \! - \! M_{K_C}(t)(M_X^n(t)-\bar{D}_n(t))}~,  
\end{align*}
due to part (ii) of Prop.~\ref{prop:mgfW}.
We now show that indeed $\bar{D}_n\equiv D_n$. 
Observe that
\begin{align*}
    \bar{D}_n(t)
    =& \sum_{k=1}^{\ceil{\alpha n}} M^{(1)}(t;k,\ceil{\beta k}) 
    M_X^{n-k}(t) \\
    & \quad \mathsf{P}(B_1 = k)\\
    =& \! \sum_{k=1}^{\ceil{\alpha n}} \!  M^{(1)}(t;k,\ceil{\beta k}) M_X^{n-k}(t) \\
    & \quad \frac{1}{{2^n}}\underset{\frac{k-1}{\alpha}<j\le\frac{k}{\alpha}}{\sum} \! \binom{n}{j},
\end{align*}
which completes the proof.
\end{proof}

\subsection{Numerical Computation of the CDF and the Tail Probability of the Completion Time}\label{sec:numericalComp}
We first describe the numerical computation of $I(t,s)$ in~\eqref{eq:I}, which is subsequently used to numerically compute $M_{W_n}(t)$.
Note that $T_{HA}$ (resp. $T_{HB}$) is given by the sum of the duration of $N_A$ (resp. $N_B$) \texttt{LINK-GEN} and $N_A-1$ (resp. $N_B-1$)\texttt{L-COMM} trails, where $N_A$ (resp. $N_B$) denotes the number trials needed for Alice (resp. Bob) to establish a link-level entanglement successfully, i.e., $N_A, N_B \overset{iid}{\sim} \text{Geo}(p_\text{gen})$.
Also, $T_{HA}$ and $T_{HB}$ are IID.
Now,
\begin{align*}
\begin{aligned}
    & I(t,s) \\
    = &\mathsf{E}\bigg(\!e^{t\max\{V_A,V_B\}} 
     e^{-\frac{|V_A-V_B|+T_{CA}+T_{CB}}{s}}\!\bigg)\\
     = & 2 e^{(t-\frac{2}{s})a_\text{com}}\left(\frac{\lambda_\text{com}}{\lambda_\text{com}\!+\!\frac{2}{s}\!-\!t}-\frac{2 \lambda_\text{com}}{2 \lambda_\text{com}\!+\!\frac{2}{s}\!-\!t}\right) \\
     & \expect{e^{(\lambda_\text{com}+\frac{1}{s}) T_{HA}} e^{-(\lambda_\text{com}+\frac{1}{s}-t)T_{HB}} \indicator{T_{HB}>T_{HA}}} \\
     & + \frac{\lambda_\text{com}}{\lambda_\text{com}\!+\!\frac{2}{s}\!-\!t} \expect{e^{\frac{1}{s}T_{HA}} e^{-(\frac{1}{s}-t)T_{HB}}\indicator{T_{HB}>T_{HA}}}.
\end{aligned}
\end{align*}
To evaluate this expectation, we derive an expression for the quantities of the form $\expect{e^{\eta T_{HA}} e^{-(\eta-t)T_{HB}}\indicator{T_{HB}>T_{HA}}}$. 
Note that $T_{HA} = \sum_{i = 1}^{N_A}T_{GA}^{(i)}+\sum_{i = 1}^{N_A-1}T_{CA}^{(i)}$, where~\footnote{Sums of the form $\sum_1^0$ are zero by convention.} $T_{GA}^{(i)} \overset{iid}{\sim}T_{GA}$ and $T_{CA}^{(i)} \overset{iid}{\sim}T_{CA}$.
We further denote $S_{GA}^{(j)} = \sum_{i = 1}^{j}T_{GA}^{(i)}$, $S_{CA}^{(j)} = \sum_{i = 1}^{j}T_{CA}^{(i)}$ (resp. for B).
Clearly, $S_{GA}^{(j)} \overset{d}{=} j a_\text{gen}+\text{Gamma}(j,\lambda_\text{gen})$, $S_{CA}^{(j)} \overset{d}{=} j a_\text{com}+\text{Gamma}(j,\lambda_\text{com})$, and $S_{GA}^{(j)}$ and $S_{CA}^{(j)}$ are independent for $j \in \mathbb{N}$ (resp. for B).
Therefore,
\begin{align}\label{eq:etaGammaHalfPlane}
\begin{aligned}
  &\expect{e^{\eta T_{HA}} e^{-(\eta-t)T_{HB}}\indicator{T_{HB}>T_{HA}}} \\
  = & p_\text{gen}^2 \sum_{l = 0}^\infty \sum_{k = 0}^\infty (1-p_\text{gen})^{k+l} \mathsf{E} \bigg(e^{\eta (S_{GA}^{(l+1)}+S_{CA}^{(l)})} \\
  & \quad  e^{-(\eta-t)(S_{GB}^{(k+1)}+S_{CB}^{(k)})} \\ & \quad \indicator{S_{GB}^{(k+1)}+S_{CB}^{(k)}>S_{GA}^{(l+1)}+S_{CA}^{(l)}} \bigg)~.
\end{aligned}    
\end{align}
We perform numerical integration to evaluate the expectation in the summand of~\eqref{eq:etaGammaHalfPlane}, which is straightforward as $S_{GA}^{(j)}$ and $S_{CA}^{(j)}$ are independent for $j \in \mathbb{N}$ and the corresponding densities are known (shifted Gamma). 
Subsequently, we compute the sum until a cut-off point (e.g., $0 \le l \le 128$, $0 \le k \le 128$).
Since we only need to evaluate $I(t,s)$ at $s = \infty$ and $s = t_\text{coh}$,~\eqref{eq:etaGammaHalfPlane} has to be evaluated only for $\eta = \lambda_\text{com}+1/t_\text{coh}, 1/t_\text{coh}, \lambda_\text{com}, 0$.

To calculate the CDF of $W_n$ according to~\eqref{eq:CDF}, we use off-the-shelf numerical inverse Laplace transform algorithms such as \texttt{invertlaplace} of the \texttt{mpmath} library of Python.
Further, we calculate the Chernoff's bound numerically as follows:
$$\mathsf{P}(W_n \! > \! s) \le \!  \min_{j \in [r]} \{e^{-ts}M_{W_n}(t_j)\}~,$$
where $M_{W_n}(t_j)<\infty$ and $r \in \mathbb{N}$, e.g., $10$.
Note that this is an approximate bound as $M_{W_n}(t_j)$'s are numerically estimated using~\eqref{eq:etaGammaHalfPlane}.
In future work, we plan to calculate the error bounds for this estimation, which would further produce an upper bound for the tail probability.

%% file: sections/sytheticSimulation.tex
\section{Synthetic Simulation of the Completion Time of the BB84 Protocol}\label{sec:synSim}

As seen in Sect.~\ref{sec:bb84duration}, successful realisation of the BB84 protocol requires repeated execution of teleportation trials, which are composed of further generation and communication phases.
This makes simulating the completion time of the protocol ($W_n$) time-consuming.
To that end, we propose an algorithm to efficiently simulate the duration $W_n$ when the distributions of the duration of individual phases follow shifted exponential distribution (assumption~\ref{A6:durationsSE}) and the respective parameters are known or can be estimated. \looseness = -1

Recall that as part of the BB84 protocol, a fraction of the teleported qubits are checked for correctness and the remaining are used as key bits, i.e., left unchecked.
The central idea of the proposed algorithm is to find the distribution of the count of unchecked qubits, qubits teleported successfully, and those teleported unsuccessfully until completion.
Subsequently, we fit a distribution to the teleportation time of a single qubit (i) unconditionally, (ii) given the teleportation attempt is successful, and (iii) given the attempt ended in a failure.
The final duration $W_n$ is then given by the sum total of (i) the aggregated unconditional teleportation time, (ii) the aggregated successful teleportation time, (iii) the aggregated failed teleportation time and (iv) the total reconciliation time.
The acceleration of our scheme is due to the fact that such aggregated duration can be simulated in a constant number of steps, irrespective of the number of qubits $n$ for an appropriate choice of distributions for the single qubit teleportation times. 
Note that this method is advantageous only when we assume that the simulation exercise will be performed for a large number of times.
In that case, distribution fitting can be done once and the output can be reused for subsequent simulation exercises, which would make this approach advantageous.
\looseness = -1

The next result states that the completion time of a failure-prone protocol can be simulated backwards if the distributions of the individual phase durations are known given the success/failure of an attempt.
The result follows simply by rearranging the terms of the MGF from~\eqref{eq:mgfW}.

\begin{lemma}\label{lemma:RUS}
Let $Y$ be a Bernoulli RV
and $X, W, T_C$ be RVs 
 such that
$$\expect{e^{tW}} = \frac{M_{T_C}(t) \expect{e^{tX} \indicator{Y = 1}}}{1-M_{T_C}(t)\expect{e^{tX} \indicator{Y = 0}}},$$
where $M_{T_C}(t) = \expect{e^{tT_C}}$ and the MGFs of $X$ and $T_C$ exist in a neighbourhood of zero.
Then, $$W \overset{d}{=} \Tilde{W} = \sum_{j=1}^{N} T_j + \sum_{j=1}^{N-1} X_j^{(0)} + X_1^{(1)},$$ 
where $T_j$'s, $X_j^{(0)}$'s, $X_j^{(1)}$'s, and $N$ are drawn independently from the following known distributions: $T_j \overset{iid}{\sim} T_C$, $X_j^{(0)} \overset{iid}{\sim} X|Y=0$, $X_j^{(1)} \overset{iid}{\sim} X|Y=1$ for $j \in \mathbb{N}$, and $N \sim \text{Geo}(\prob{Y=1})$.
\end{lemma}
\begin{proof}
 See Sect.~\ref{sec:defProofs}.
\end{proof}

Let us now recall a few features of the BB84 protocol from Sect.~\ref{sec:bb84duration}: it is deemed successful when at least one qubit is sampled, and out of the sampled qubits, a certain fraction ($\beta$) results in faithful teleportation.
Reusing notations from Lemma~\ref{lemma:RUS} and Sect.~\ref{sec:bb84duration}, let $N$ denote the number of times the protocol has to be run to see the first success.
That is,
\begin{align*}
  &N \sim \text{Geo}(\mathsf{P}(U=1))~, \quad \text{where} \\
  & U = \mathbbm{1}_{\sum_{i=1}^{B_1}\! Y_{k_i} \ge \beta B_1}\mathbbm{1}_{B_1 \ge 1}~,
\end{align*}
and $k_i$'s for $i \in [B_1]$ denote the exhaustive set of indices for which $Z_{k_i} = 1$.
Applying Lemma~\ref{lemma:RUS}, \eqref{eq:mgfWn} gives:
\begin{align} \label{eq:simWn1}
  W_n \overset{d}{=} \Tilde{W}_n = \sum_{j=1}^{N} T_j + \sum_{j=1}^{N-1}  V_{j}^{(0)} + V_1^{(1)}~,
\end{align}
where $T_j \overset{iid}{\sim} T_C$, $V_j^{(l)} \overset{iid}{\sim} \sum_{i=1}^n X_i|U\!=\!l$, for $l \in \{0,1\}$ and $j \in \mathbb{N}$, and $N \sim \text{Geo}(\prob{U=1})$.

Now for a single BB84 trial, let $N_S$ and $N_F$ respectively denote the number of qubits out of the sampled ones that were faithfully teleported and those ended in a failure, i.e., 
$$N_S = \sum_{i=1}^{B_1}\! Y_{k_i}~, \quad N_F = \sum_{i=1}^{B_1}\! (1-Y_{k_i})~.$$
The following result provides a way to simulate $V_j^{(0)}$'s and $V_j^{(1)}$'s.

\begin{prop}\label{prop:synSim}
Let $(N_S^{(l)},N_F^{(l)}) \sim (N_S,N_F)|U = l$ for $l \in \{0,1\}$.
 Then,
 $$V_j^{(l)} \!\overset{d}{=} \Tilde{V}_l \!=\! \sum_{i=1}^{N_S^{(l)}} X_i^{(1)} + \sum_{i=1}^{N_F^{(l)}} X_i^{(0)} +\! \! \sum_{i=1}^{n-N_S^{(l)}-N_F^{(l)}} \!\! X_i~,$$
where $X_j^{(l)} \overset{iid}{\sim} X_1|Y=l$, $l \in \{0,1\}$ and all summands on the RHS are drawn independently of each other and of  $(N_S^{(l)}, N_F^{(l)})$ as well.
\end{prop}
\begin{proof}
 See Sect.~\ref{sec:defProofs}.
\end{proof}
We can now use~\eqref{eq:simWn1} together with Prop.~\ref{prop:synSim} to simulate $W_n$, the completion time of the protocol, provided we can simulate (i) the number of protocol runs $N$, (ii) the conditional counts of teleportation successes and failures in a trial $(N_S, N_F)|U$, and (iii) the single qubit teleportation durations distributed as $X$, $X|Y = 1$, and $X|Y = 0$.
For simulating $N$, we observe that \looseness = -1
\begin{align*}
    & P(U=1)\\
    =& \sum_{k=1}^{\ceil{\alpha n}} \mathsf{P}\bigg(\sum_{j=1}^{k}\! Y_{i_j} \ge \beta k \bigg)
    \mathsf{P}(B_1 = k)\\
    =& \! \sum_{k=1}^{\ceil{\alpha n}} \!  \mathsf{P}\bigg(\!\text{Bin}(k,\!m_1(0)) \!\ge\!\beta k \!\bigg)\frac{1}{2^n} \!  \underset{\frac{k-1}{\alpha}<j\le\frac{k}{\alpha}}{\sum} \! \binom{n}{j},
\end{align*}
where we have used 
the fact that $Y_i$'s are IID Bernoulli with $\mathsf{P}(Y_i\!=\!1)\!=\!m_1(0)$. 
Denoting $P(U=1) \!=\! p_1$,
\begin{align}\label{eq:sfu0}
    &\mathsf{P}( N_S = s,N_F = f|U=0 ) \nonumber\\
    =& (1-\mathbbm{1}_{\frac{s}{s+f} \ge \beta}) \thickspace   \mathsf{P}\bigg(\!\text{Bin}(s+f,\!m_1(0)) \!=\!s \!\bigg)\nonumber\\
    & \quad \frac{1}{2^n} \!  \underset{\frac{s+f-1}{\alpha}<j\le\frac{s+f}{\alpha}}{\sum} \! \binom{n}{j} \quad  (1-p_1)^{-1}.
\end{align}
Similarly,
\begin{align}\label{eq:sfu1}
    &\mathsf{P}( N_S = s,N_F = f|U=1 ) \nonumber\\
    =&  \mathsf{P}( N_S = s,N_F = f,U=1 ) p_1^{-1}\nonumber\\
    =& \mathbbm{1}_{\frac{s}{s+f} \ge \beta} \thickspace \mathsf{P}\bigg(\!\text{Bin}(s+f,\!m_1(0)) \!=\!s \!\bigg)\nonumber\\
    & \quad \frac{1}{2^n} \!  \underset{\frac{s+f-1}{\alpha}<j\le\frac{s+f}{\alpha}}{\sum} \! \binom{n}{j} \quad p_1^{-1}~.
\end{align}

Finally, we need a way to simulate from the distributions of the single qubit teleportation times: $X$, $X|Y = 1$, and $X|Y = 0$.
We approximate them by three shifted Coxian phase-type distributions using the method of moments.
Note that the moments of the qubit teleportation times can be calculated using~\eqref{eq:m1} and~\eqref{eq:m0Mx}. 
The same equations also help us determine the shifts, which turn out to be $a_W = a_{\text{gen}}+a_{\text{com}}+a_{\text{swap}}+a_{\text{AB}}$ for all three distributions.
Since a Coxian phase-type distribution with $d$ phases has $(2d-1)$ parameters, we can compute the first $(2d-1)$ moments as functions of the parameters and solve for the parameters by equating them to the moments derived from~\eqref{eq:m1} and~\eqref{eq:m0Mx}.  
Note that the distribution fitting is done only once and the output can be reused for subsequent simulation exercises.

\SetKwComment{Comment}{/* }{ */}
\begin{algorithm}
\caption{Synthetic Simulation of the completion time of the BB84 protocol}\label{alg:two}
\small{
\KwData{$n \in \mathbb{N}$, $\alpha>0$, $\beta>0$, $a_\text{gen}>0$, $a_\text{com}>0$, $a_\text{swap}>0$, $a_\text{AB}>0$, $\lambda_\text{gen}>0$, $\lambda_\text{com}>0$, $\lambda_\text{swap}>0$, $\lambda_\text{AB}>0$, $0<p_\text{gen}<1$, $0<p_\text{swap}<1$, $t_c>0$, $t_{de}>0$, $t_{da}>0$, $d \in \mathbb{N}$.}
\KwResult{$W_n$}
\textbf{Preprocessing}: \\
$a_W \gets a_{\text{gen}}+a_{\text{com}}+a_{\text{swap}}+a_{\text{AB}}$
$p_1 \! \gets \! \sum_{k=1}^{\ceil{\alpha n}} \!  \mathsf{P}\bigg(\!\text{Bin}(k,\!m_1(0)) \!\ge\!\beta k \!\bigg)\frac{1}{2^n} \!  \underset{\frac{k-1}{\alpha}<j\le\frac{k}{\alpha}}{\sum} \! \binom{n}{j}$\;
Estimate $(\lambda_1^{(1)}, \dots, \lambda_d^{(1)}, q_1^{(1)}, \dots,q_{d-1}^{(1)})$ for $X|Y=1$ via moment-matching\;
Estimate $(\lambda_1^{(0)}, \dots, \lambda_d^{(0)}, q_1^{(0)}, \dots,q_{d-1}^{(0)})$ for $X|Y=0$ via moment-matching\;
Estimate $(\lambda_1^{(u)}, \dots, \lambda_d^{(u)}, q_1^{(u)}, \dots,q_{d-1}^{(u)})$ for $X$ via moment-matching\;
\textbf{Simulation}: \\
Draw $N \sim \text{Geo}(p_1)$\;
Draw $(S_0,F_0)$ according to~\eqref{eq:sfu1}\;
\For{$i = 1:N-1$}{
  Draw $(S_i,F_i)$ according to~\eqref{eq:sfu0};
}
$S \gets \sum_{j = 0}^{N-1}S_i$\;
$F \gets \sum_{j = 0}^{N-1}F_i$\;
$\Tilde{N} \gets nN-S-F$\;
$A_0 \gets S$\;
Draw $W_A^{(0)} \sim \text{Gamma}(A_0,\lambda_1^{(1)})$\;
$B_0 \gets F$\;
Draw $W_B^{(0)} \sim \text{Gamma}(B_0,\lambda_1^{(0)})$\;
$C_0 \gets \Tilde{N}$\;
Draw $W_C^{(0)} \sim \text{Gamma}(C_0,\lambda_1^{(u)})$\;
\For{$i = 1:d-1$}{
  Draw $A_i \sim \text{Bin}(A_{i-1},q_i^{(1)})$\;
  Draw $W_A^{(i)} \sim \text{Gamma}(A_i,\lambda_{i+1}^{(1)})$\;
  Draw $B_i \sim \text{Bin}(B_{i-1},q_i^{(0)})$\;
  Draw $W_B^{(i)} \sim \text{Gamma}(B_i,\lambda_{i+1}^{(0)})$\;
  Draw $C_i \sim \text{Bin}(C_{i-1},q_i^{(u)})$\;
  Draw $W_C^{(i)} \sim \text{Gamma}(C_i,\lambda_{i+1}^{(u)})$\;
}
Draw $C \sim \text{Gamma}(N,\lambda_{AB})$
$W_n \gets N a_{AB}+C+nNa_W+\sum_{i = 0}^{d-1}(W_A^{(i)}+W_B^{(i)}+W_C^{(i)}).$
}
\end{algorithm}

The choice of Coxian phase-type distribution is motivated by the fact that their IID sum can be expressed in a compact form and thus can be efficiently simulated.
Note that the sum of $k$ IID Coxian phase-type RVs with parameters $(\lambda_1, \dots, \lambda_d, q_1, \dots,q_{d-1})$ is distributed as $\sum_{l=1}^{d} \text{Gamma}(D_l,\lambda_l)$ where $D_l \sim \text{Bin}(D_{l-1},q_{l-1})$ for $l \ge 2$ and $D_1 = k$.
The efficiency of our approach follows from the fact that, on average, it needs to simulate $(2d\!+\!1\!+\!\frac{2}{p_1})$ RVs per observation vis-à-vis $(n((\frac{2+4(1-p_{\text{gen}})}{1-(1-p_{\text{gen}})^2}\!+\!1)\frac{1}{p_{\text{swap}}}\!+\!1)\!+\!1)\frac{1}{p_1}$ required for the full-scale approach.
Recall that $p_1 = \prob{U=1}$, $(1-p_\text{swap})$ denotes the observable BSM failure probability at the repeater, and $p_\text{gen}$ denotes the link-level entanglement generation success probability.

\begin{figure*}[t]
\centering
\begin{subfigure}{0.32\textwidth}
    \centering
    \includegraphics[width=1\textwidth]{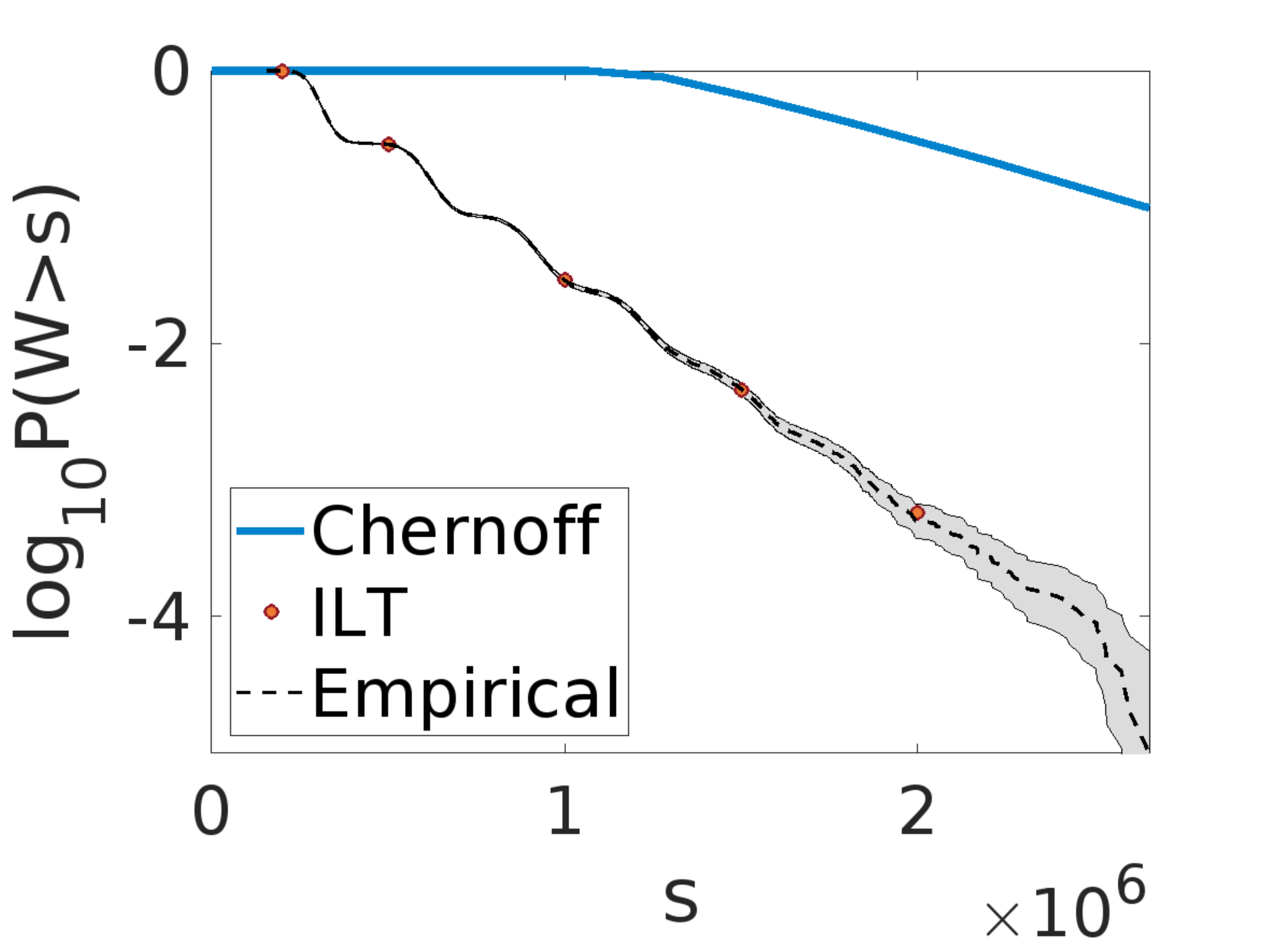}
    \caption{\label{fig:bb84_1}%
    $p_\text{gen} =  10^{-3}$}
\end{subfigure}
    \hfill
\begin{subfigure}{0.32\textwidth}
    \centering
    \includegraphics[width=1\textwidth]{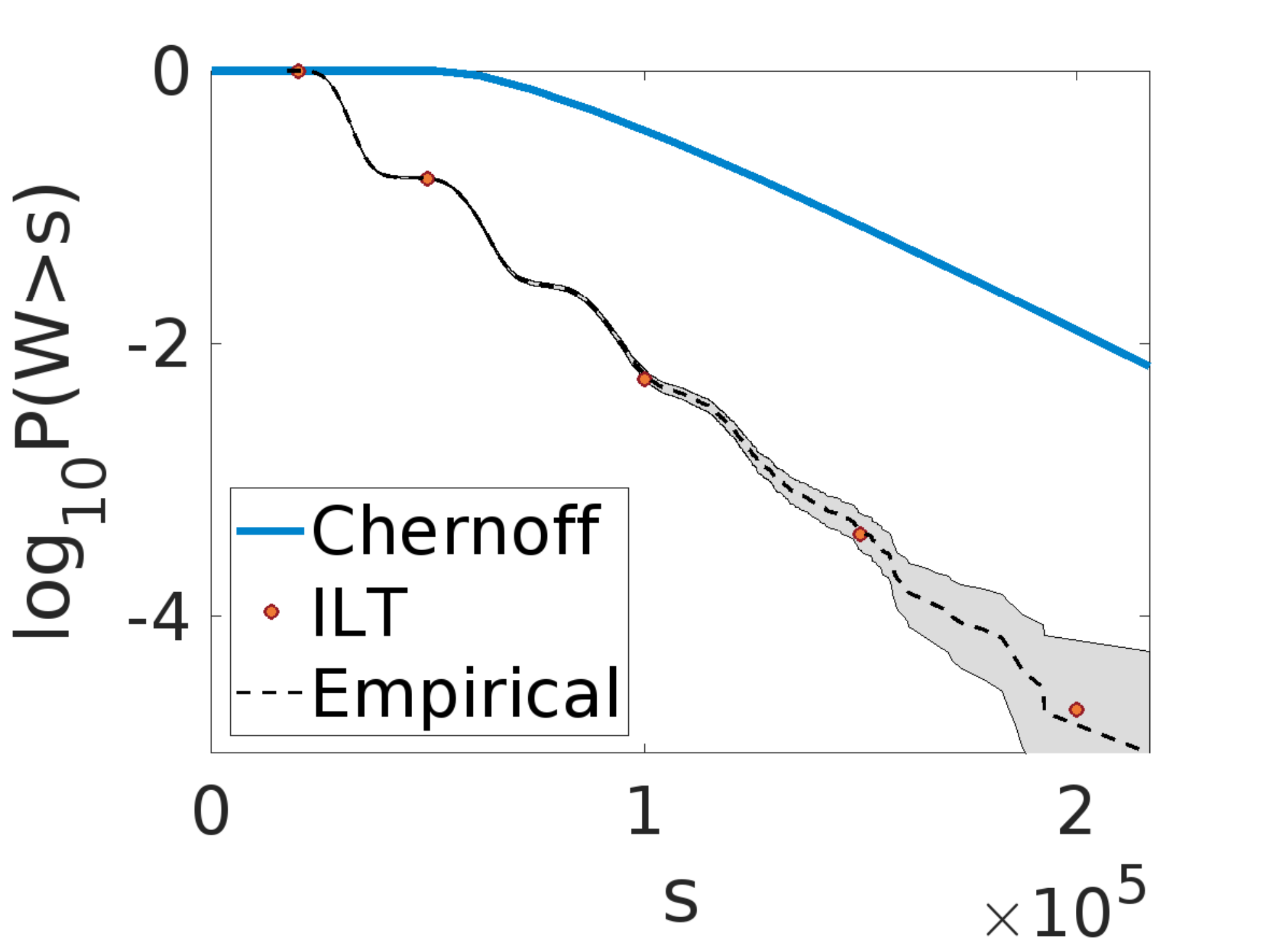}
    \caption{\label{fig:bb84_2}%
    $p_\text{gen} =  10^{-2}$}
\end{subfigure}
\hfill
\begin{subfigure}{0.32\textwidth}
    \centering
    \includegraphics[width=1\textwidth]{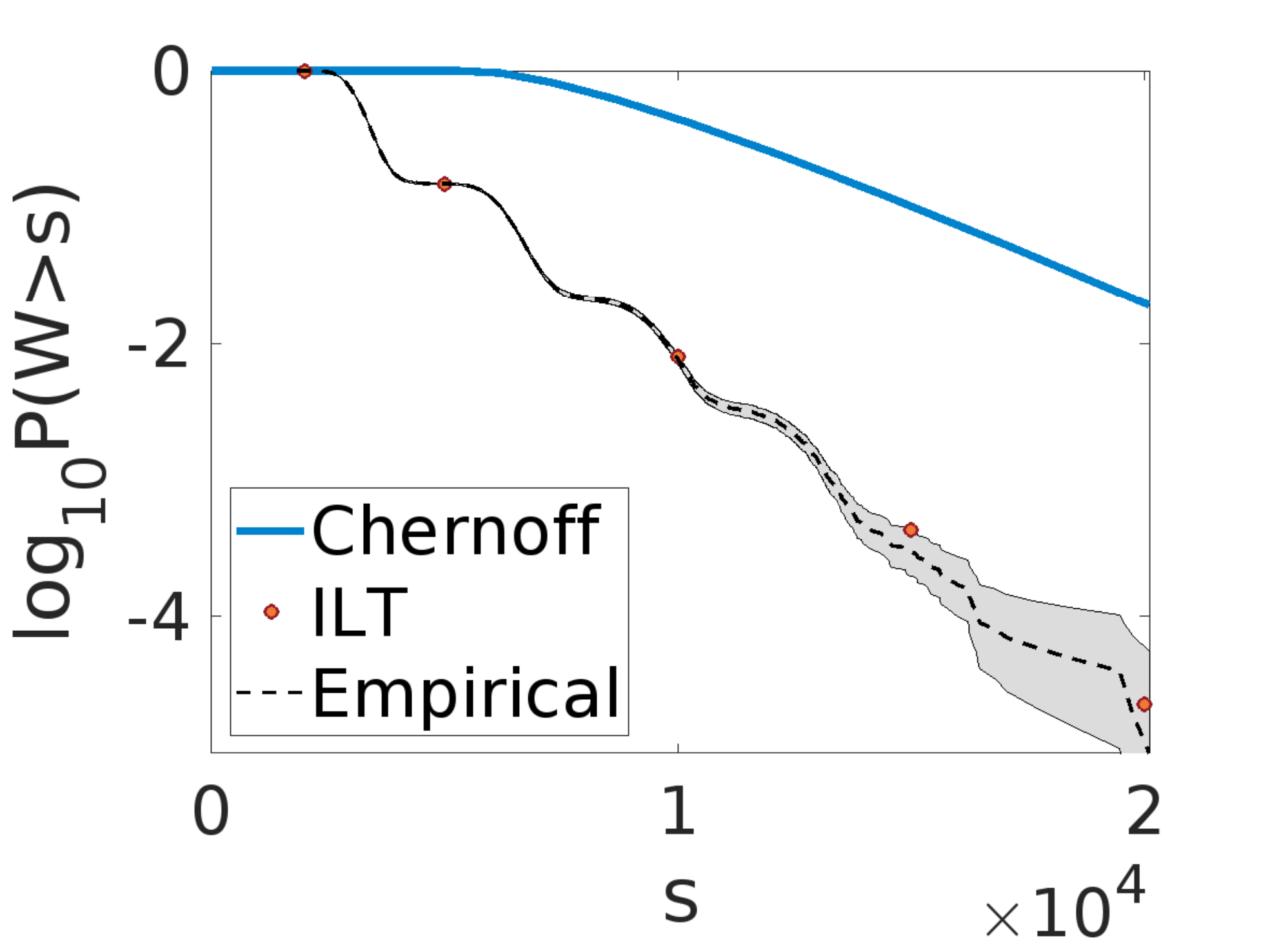}
    \caption{\label{fig:bb84_3}%
    $p_\text{gen} =  10^{-1}$}
\end{subfigure}%
\caption{\label{fig:bb84chf}%
Numerical estimates of the CDF via inverse Laplace transform (`ILT') and the Chernoff's bound (`Chernoff') vis-\`{a}-vis the empirical tail probability (`Empirical') of the completion time ($W_n$) of the BB84 protocol for $n=50$ qubits; the shaded region represents $95\%$ confidence interval for the CCDF computed according to~\cite[Thm. 2.4]{jylbPerf}. The empirical distribution is obtained by running a full-scale simulation; see Sect.~\ref{sec:eval} for an exhaustive list of parameters of the simulation.  Time unit: mean link-level entanglement generation time.}%
\end{figure*}

\begin{figure}[t]
    \centering \includegraphics[width=0.95\linewidth]{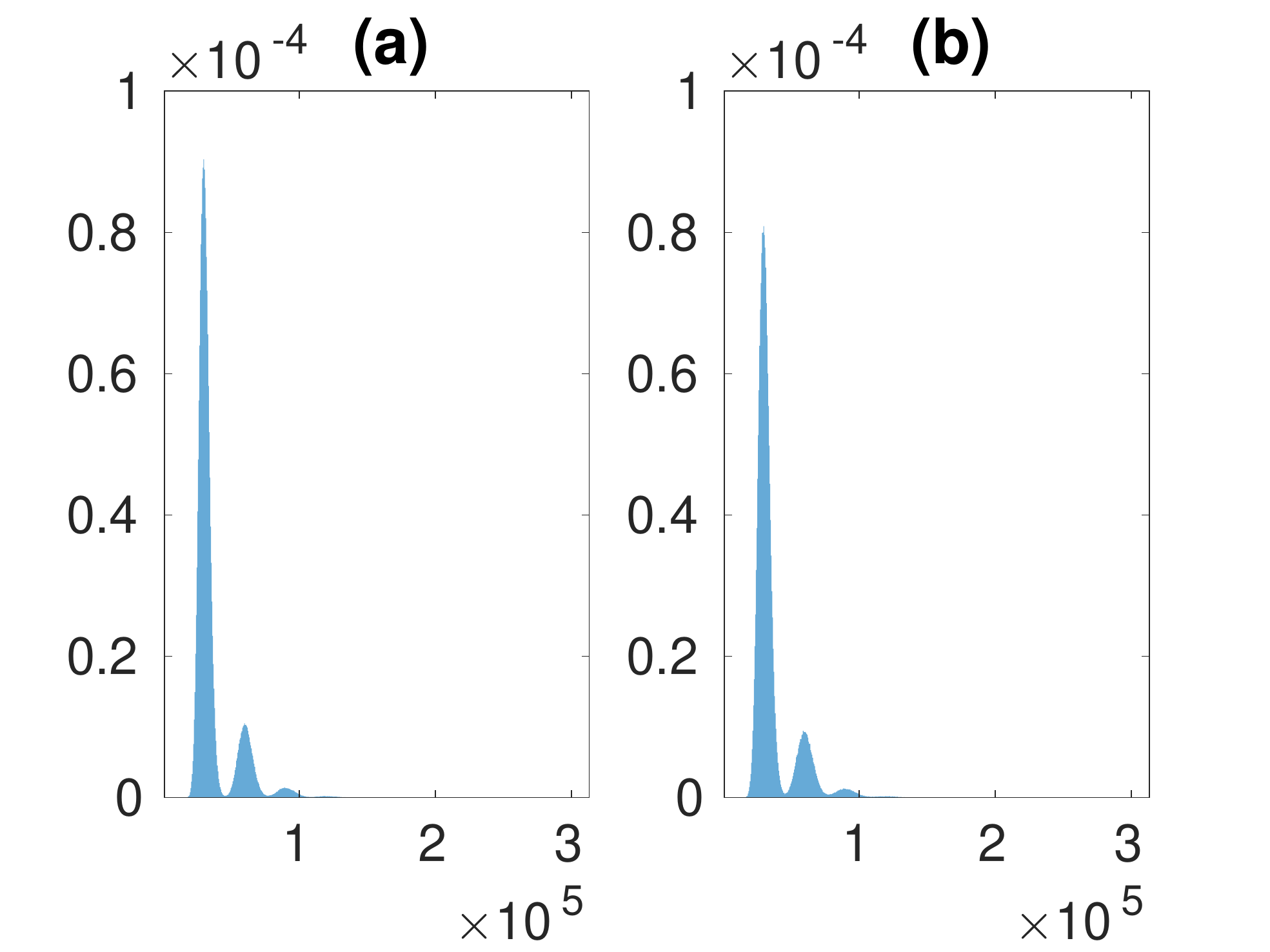}
    \caption{Comparison of the empirical densities of the protocol completion time $W_n$ from a (\textbf{a}) full-scale and a (\textbf{b}) synthetic simulation. Parameters: $n = 50$ qubits; link generation success probability $p_{\text{gen}} = 10^{-2}$; other parameters are provided in Sect.~\ref{sec:eval}.}
    \label{fig:histCompare}
\end{figure}

\begin{figure}[t]
    \centering \includegraphics[width=0.95\linewidth]{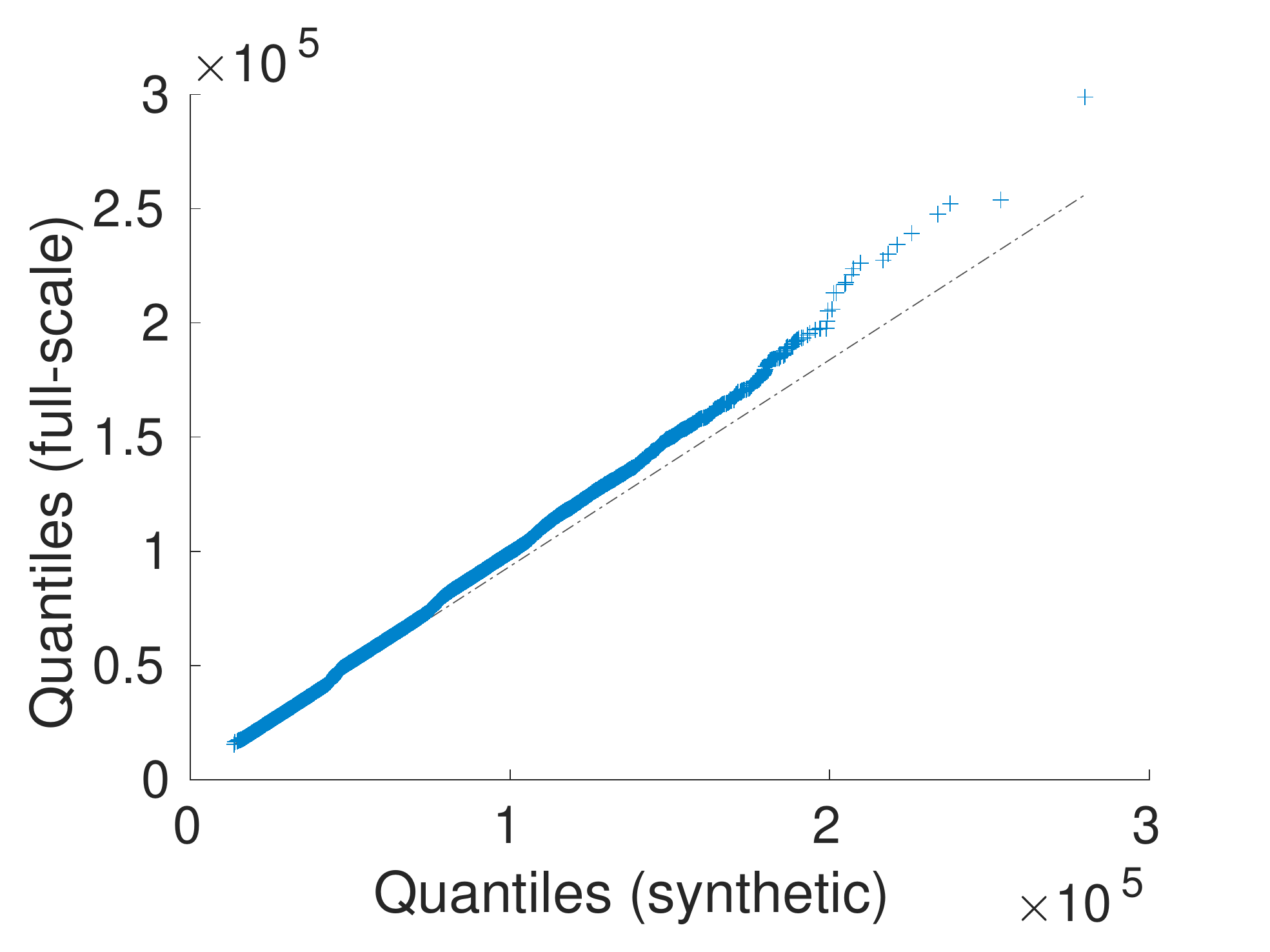}
    \caption{Q-Q plot of the protocol completion times from a full-scale and a synthetic simulation; the experimental setup is identical to Fig.~\ref{fig:histCompare}.}
    \label{fig:qqPlot}
\end{figure}

The complete simulation scheme is formalised in Algo.~\ref{alg:two}.
We also compare our approach to a full-scale simulation based on $10^6$ observations in (i) Fig.~\ref{fig:histCompare}, where the histograms of the outputs from the two approaches are compared, and in (ii) Fig.~\ref{fig:qqPlot}, where a QQ-plot of these outputs is presented.
In the synthetic approach, we use Coxian phase-type distributions with $7$ phases as they turn out to be reasonably good fits.
The exhaustive list of parameters used for the simulations can be found in Sect.\ref{sec:eval}.

%% file: sections/eval.tex
\section{Numerical Evaluations}\label{sec:eval}

In this section, we evaluate the estimate of the CDF and the tail bound derived earlier in Sect.\ref{sec:bb84duration} 
numerically.
The time unit across plots is the mean link-level entanglement generation time.
In the context of the completion time of the BB84 protocol, we first run a full-scale simulation and compare the bound/estimate with the empirical CCDF.
The comparison is shown in Fig.~\ref{fig:bb84chf} for three different values of the entanglement generation probability $\{10^{-1},10^{-2},10^{-3}\}$ to reflect different levels of hardware efficiency.
In the experiments, we assume that the phases \texttt{LINK-GEN}, \texttt{L-COMM}, and \texttt{S-COMM} follow $\text{SE}(\frac{1}{2},2)$ distribution, i.e., $a_\text{gen}\!=\!a_\text{com}\!=\!a_\text{swap}\!=\!\frac{1}{2}$ and $\lambda_\text{gen}\!=\!\lambda_\text{com}\!=\!\lambda_\text{swap}\!=\!2$, while the phases \texttt{T-COMM} and \texttt{K-COMM} are assumed to follow $\text{SE}(1,1)$ distribution, i.e., 
$a_\text{AB}\!=\!1$ and $\lambda_\text{AB}\!=\!1$.
The Werner parameter for a fresh link is assumed to be $w_0 \!=\! 0.98$.
Further, the constants characterising memory decoherence are taken to be $t_c \!=\! t_{de} \!=\! t_{da} \!=\! 4 \times 10^4$.
We run the BB84 protocol for $n = 50$ qubits, while $\alpha = 30\%$ of the qubits with the same measurement bases are chosen for reconciliation.
The protocol is deemed successful if we sample at least one qubit for reconciliation and the measurement results agree for at least $\beta = 95\%$ cases.
In the resulting plots in Fig.~\ref{fig:bb84chf}, we see that our estimate of the CDF matches closely with the empirical CDF for sampled points.

Next, we demonstrate the effectiveness of our simulation scheme in Sect.~\ref{sec:synSim}.
We compare the protocol completion times generated from our scheme vis-\'a-vis a full-scale simulation for a system with entanglement generation probability of $10^{-2}$ while keeping other factors the same as in Fig.~\ref{fig:bb84chf}.
We compare the resulting empirical densities in Fig.~\ref{fig:histCompare}, whereas a Q-Q plot is presented in Fig.~\ref{fig:qqPlot}.
We observe good agreement between the two approaches.

%% file: sections/conclusion.tex
\section{Conclusion}\label{sec:conclusion}

In this work, we did a performance analysis of the completion time of the BB84 protocol.
Our setup assumes that the sender and the receiver are connected by a single quantum repeater and that there is no eavesdropping in the quantum channel between them. 
To reflect the current quantum hardware standards, we however consider the possibility of failure at every individual phase of teleportation and take into account the resulting effect of decoherence on the performance of the protocol.
We subsequently provide a method to calculate the MGF of the completion time, which lets us calculate an estimate of the CDF via inverse Laplace transform and a numerical bound for the corresponding tail probability.
Making certain assumptions on the distributions of the durations of individual phases, we also propose an efficient simulation scheme for the completion time.
The simulation scheme is based on the idea that the single qubit teleportation time can be well-approximated by a Coxian phase-type distribution, which can be efficiently aggregated to arrive at the completion time of the whole protocol. \looseness = -1

%% file: sections/appendix.tex
\section*{Appendix}\label{sec:appendix}

\subsection{Deferred Proofs}\label{sec:defProofs} 

\begin{proof}[Proof of Prop.~\ref{prop:existenceMgf}]
For $t \le 0$,
\begin{align*}
    e^{t \max_{i \in [n]}T_i} p_{\theta}(\mathbf{L}_1) & \le e^{t T_1}~.
\end{align*}
Similarly, for $t>0$,
\begin{align*}
  e^{t \max_{i \in [n]}T_i} p_{\theta}(\mathbf{L}_1) & \le e^{t \sum_{j=1}^n T_j}~.
\end{align*}
Since $M_{T_C}(t) \ge 0$, for $t \in \mathbb{R}$ we have:
\begin{align*}
   & M_{T_C}(t) \mathsf{E}\big(e^{t \max_{i \in [n]}T_i} p_{\theta}(\mathbf{L}_1)\big)\\
   \le & M_{T_C}(t) \max \big\{\mathsf{E}(e^{t T_1}), \prod_{j=1}^n \mathsf{E}(e^{t T_j})\big\}~.
\end{align*}
The RHS is finite on $\bigcap_{l \in [n]\cup\{c\}} I_l$.
The same holds for the denominator of the RHS of~\eqref{eq:mgfW}.
Further, the infinite sum on the second line of~\eqref{eq:mgfW} converges on $I_0$, which proves the claim.
\end{proof}

\begin{proof}[Proof of Prop.~\ref{prop:subExp}]
Recall that an RV is sub-exponential iff its MGF exists in a neighbourhood of zero.
Thus, given the hypothesis, it is enough to show that $I_0$ is non-empty.
For $t_1<t_2$ and $t_1, t_2\in \cap_{i \in [n] \cup \{C\}} I_l$,
\begin{align*}
    &G(t_2)-G(t_1)\\ 
    = & \mathsf{E}\big((e^{t_2 X_1}-e^{t_1 X_1}) (1-p_{\theta}(\mathbf{L}_1))\big) \\
    \le & \mathsf{E}(e^{t_2 X_1})-\mathsf{E}(e^{t_1 X_1})~.
\end{align*}
Thus, continuity of the MGF of $X_1$ in $\cap_{i \in [n]\cup \{C\}} I_l$ implies continuity of $G$.
Further, barring the trivial case $p_{\theta}(\mathbf{L}_1) \!\equiv\! 0$, $G(0) \!=\! \mathsf{E}(1\!-\!p_{\theta}(\mathbf{L}_1))\!<\!1$.
Since $G$ is continuous, $I_0$ is non-empty as claimed.
\end{proof}



\begin{proof} [Proof of Lemma~\ref{lemma:RUS}]
Denoting $p_l = P(Y = l)$ and $\phi_l(t) = \expect{e^{t X_1^{(l)}}} = \expect{e^{tX} \indicator{Y = l}}/p_l$ for $l \in \{0,1\}$,
\begin{align*}
    &\expect{\!e^{t \Tilde{W}}} \\
    = & M_{T_C}(t) \phi_1(t) \sum_{j = 0}^{\infty}  p_0^j p_1 (M_{T_C}(t) \phi_0(t))^j \\
    = &\frac{p_1 M_{T_C}(t) \phi_1(t)}{1-p_0 M_{T_C}(t) \phi_0(t)} \\
    = &\expect{e^{t W}}~.
\end{align*}
Following the argument of Corr.~\ref{prop:subExp}, MGFs of $W$ and $\Tilde{W}$ exist in a neighbourhood of zero, implying that $W \overset{d}{=} \Tilde{W}$.
\end{proof}

\begin{proof}[Proof of Prop.~\ref{prop:synSim}]
Observe that
\begin{align*}
    &\expect{e^{t V_j^{(l)}}}\\
    = & \expect{e^{t \sum_{1}^n \!X_i}| U \!=\! l} \\
    = &\sum_{k,d}\mathsf{E}\big(e^{t \sum_{1}^n \!X_i}|\sum_{1}^n Y_i Z_i \!=\! k, \sum_{1}^n Z_i \!=\! d, U \!=\! l \big) \\
    & \quad \mathsf{P}\big(\sum_{1}^n Y_i Z_i \!=\! k, \sum_{1}^n Z_i \!=\! d| U \!=\! l \big) \\
    = &\sum_{k,d}\mathsf{E}\big(e^{t \sum_{1}^n \!X_i}|\sum_{1}^n Y_i Z_i \!=\! k, \sum_{1}^n Z_i \!=\! d \big) \\
    & \quad \mathsf{P}\big(\sum_{1}^n Y_i Z_i \!=\! k, \sum_{1}^n Z_i \!=\! d| U \!=\! l \big)~,
\end{align*}
where the last equality follows from the definition of $U$.
Now, for $X \sim X_1$ and $Y \sim Y_1$,
\begin{align*}
    &\mathsf{E}\big(e^{t \sum_{1}^n \!X_i}|\sum_{1}^n Y_i Z_i \!=\! k, \sum_{1}^n Z_i \!=\! d \big) \\
    = & \sum_{r = k}^{k+n-d}\mathsf{E}\big(e^{t \sum_{1}^n \!X_i}| \sum_{1}^n Y_i \!=\! r, \sum_{1}^n Y_i Z_i \!=\! k, \\
    & \quad \sum_{1}^n Z_i \!=\! d \big) \mathsf{P}\big(\sum_{1}^n Y_i \!=\! r | \sum_{1}^n Y_i Z_i \!=\! k, \\
    & \quad \sum_{1}^n Z_i \!=\! d \big) \\
    = & \sum_{r = k}^{k+n-d}\mathsf{E}\big(e^{t \sum_{1}^n \!X_i}| \sum_{1}^n Y_i \!=\! r \big) \\
    & \quad \mathsf{P}\big(\sum_{1}^n Y_i \!=\! r | \sum_{1}^n Y_i Z_i \!=\! k, \sum_{1}^n Z_i \!=\! d \big) \\
     = & \mathsf{E}^k\big(e^{t X}| Y \!=\! 1 \big) \mathsf{E}^{d-k}\big(e^{t X}| Y \!=\! 0 \big) \\
     & \quad \sum_{r = 0}^{n-d} \binom{n-d}{r}\mathsf{E}^r\big(e^{t X}| Y \!=\! 1 \big) (\prob{Y=1})^r \\
     & \quad \mathsf{E}^{n-d-r}\big(e^{t X}| Y \!=\! 0 \big) (\prob{Y=0})^{n-d-r} \\
     = & \mathsf{E}^k\big(e^{t X}| Y \!=\! 1 \big) \mathsf{E}^{d-k}\big(e^{t X}| Y \!=\! 0 \big) \mathsf{E}^{n-d}\big(e^{t X}\big). 
\end{align*}
Therefore,
\begin{align*}
    &\expect{e^{t V_j^{(l)}}}\\
    = &\sum_{k,d}\mathsf{E}^k\big(e^{t X}| Y \!=\! 1 \big) \mathsf{E}^{d-k}\big(e^{t X}| Y \!=\! 0 \big) \\
    & \quad \mathsf{E}^{n-d}\big(e^{t X}\big) \mathsf{P}\big(\sum_{1}^n Y_i Z_i \!=\! k, \sum_{1}^n Z_i \!=\! d| U \!=\! l \big) \\
    = &\sum_{k,d}\mathsf{E}^k\big(e^{t X}| Y \!=\! 1 \big) \mathsf{E}^{d}\big(e^{t X}| Y \!=\! 0 \big) \\
    & \quad \mathsf{E}^{n-d-k}\big(e^{t X}\big) \mathsf{P}\big(N_S \!=\! k, N_F \!=\! d| U \!=\! l \big)~,
\end{align*}
which establishes $V_j^{(l)} \!\overset{d}{=} \Tilde{V}_l$ for $l \in \{0,1\}$.
\end{proof}

%% file: tqe.bbl
\begin{thebibliography}{00}

\bibitem{vanMeterBook}
Van Meter, R., "Quantum networking", \emph{ John Wiley \& Sons}, 2014.

\bibitem{munroRepeaters}
Munro, W.J., Azuma, K., Tamaki, K. and Nemoto, K., "Inside quantum repeaters", \emph{IEEE Journal of Selected Topics in Quantum Electronics}, 21(3), pp.78-90, 2015.

\bibitem{brand}
Brand, S., Coopmans, T. and Elkouss, D., "Efficient computation of the waiting time and fidelity in quantum repeater chains",~\emph{IEEE Journal on Selected Areas in Communications}, 38(3), pp.619-639, 2020.

\bibitem{liRus}
Li, B., Coopmans, T. and Elkouss, D., "Efficient optimization of cut-offs in quantum repeater chains",~\emph{ In 2020 IEEE International Conference on Quantum Computing and Engineering (QCE)}, pp. 158-168, 2020.

\bibitem{coopmans2022improved} Coopmans, T., Brand, S. and Elkouss, D., "Improved analytical bounds on delivery times of long-distance entanglement",~\emph{Physical Review A}, vol. 105, no. 1, pp. 012608, 2022.

\bibitem{bb84qkd} 
Bennett C.H. and Brassard G., "Quantum cryptography: Public key distribution and coin tossing",~\emph{International Conference on Computers, Systems \& Signal Processing}, pp.175-179, 1984.

\bibitem{bahadur1960deviations} 
Bahadur, R.R. and Ranga Rao, R., "On deviations of the sample mean",~\emph{The Annals of Mathematical Statistics}, 31(4), pp.1015-1027, 1960.

\bibitem{qiaoSim} 
Qiao, H. and Chen, X.Y., "Simulation of BB84 Quantum Key Distribution in depolarizing channel",~\emph{In Proceedings of 14th Youth Conference on Communication}, (pp. 123-129), 2009. 

\bibitem{sahooSim}
Sahoo, J.R. and Satapathy, S., "Simulation and analysis of BB84 protocol by model checking",~\emph{International Journal of Engineering Science and Technology (IJEST)}, 3(7), 2011.

\bibitem{halipSim}
Halip, N.H.M., Mokhtar, M. and Buhari, A., "Simulation of Bennet and Brassard 84 protocol with Eve's attacks",~\emph{In 2014 IEEE 5th International Conference on Photonics (ICP)}, (pp. 29-31). IEEE, 2014.

\bibitem{jasimSim} 
Jasim, O.K., Abbas, S., El-Horbaty, E.S.M. and Salem, A.B.M., "Quantum key distribution: simulation and characterizations",~\emph{Procedia Computer Science}, 65, pp.701-710, 2015.

\bibitem{minaScalableSim} 
Mina, M.Z. and Simion, E., "A scalable simulation of the BB84 protocol involving eavesdropping",~\emph{In Innovative Security Solutions for Information Technology and Communications: 13th International Conference, SecITC} Bucharest, Romania, November 19–20, Revised Selected Papers 13 (pp. 91-109). Springer International Publishing, 2021.

\bibitem{werner}
Werner, R.F., "Quantum states with Einstein-Podolsky-Rosen correlations admitting a hidden-variable model", Physical Review A, 40(8), p.4277, 1989.

\bibitem{nielsenChuang}
Nielsen, M.A. and Chuang, I., "Quantum computation and quantum information", 2002.

\bibitem{vardoyanPerf}
Vardoyan, G., Skrzypczyk, M. and Wehner, S., "On the quantum performance evaluation of two distributed quantum architectures",~\emph{ACM SIGMETRICS Performance Evaluation Review}, 49(3), pp.30-31, 2022.

\bibitem{jylbPerf}
Le Boudec, J.Y., "Performance evaluation of computer and communication systems",~\emph{Epfl Press}, 2010.






\end{thebibliography}
